\documentclass[aps,prl,showpacs]{revtex4}

\usepackage{amsmath,amsfonts,graphicx,epsfig}

\begin{document}

\title{One-Particle and Few-Particle Billiards}

\author{Steven Lansel}
\affiliation{School of Electrical and Computer Engineering and School of
Mathematics, Georgia Institute of Technology, Atlanta, GA 30332-0160}

\author{Mason A. Porter}
\affiliation{Department of Physics and Center for the Physics of Information \\
California Institute of Technology, Pasadena, CA  91125}

\author{Leonid A. Bunimovich}
\affiliation{School of Mathematics and Center for Nonlinear Science, Georgia
Institute of Technology, Atlanta, GA 30332-0160}

\begin{abstract} 

We study the dynamics of one-particle and few-particle billiard systems in containers of various shapes.  In few-particle systems, the particles collide elastically both against the boundary and against each other.  In the one-particle case, we investigate the formation and destruction of resonance islands in (generalized) mushroom billiards, which are a recently discovered class of Hamiltonian systems with mixed regular-chaotic dynamics.  In the few-particle case, we compare the dynamics in container geometries whose counterpart one-particle billiards are integrable, chaotic, and mixed.  One of our findings is that two-, three-, and four-particle billiards confined to containers with integrable one-particle counterparts inherit some integrals of motion and exhibit a regular partition of phase space into ergodic components of positive measure.  Therefore, the shape of a container matters not only for noninteracting particles but also for interacting particles.

\end{abstract}

\date{\today}

\maketitle

\vspace{2 mm}



{\bf In this paper, we conduct a numerical investigation of one-particle systems (billiards) with regular, chaotic, and mixed (regular-chaotic) dynamics and of small numbers (two, three, and four) of elastically colliding particles (balls) confined to the same billiard tables.  Thus, this report differs essentially from traditional numerical studies of systems of interacting particles, which deal with many particles confined to containers of the simplest shapes (usually boxes or boxes with periodic boundary conditions [i.e., tori]).  Using the simple example of hard balls in a circle, we demonstrate that systems of interacting particles need not be ergodic.  Therefore, the recently-discovered typical inhomogeneity of stationary distributions of noninteracting particles in containers may well be in effect for interacting particles as well.  We also compare the dynamics of few-particle systems with their one-particle counterparts when the dynamics of the latter are regular, chaotic, and mixed.}

\section{Introduction}

Two major 20th century discoveries completely transformed scientists' understanding of nonlinear phenomena.  One was Kolmogorov-Arnold-Moser (KAM) theory, which demonstrated the stability of regular dynamics for small perturbations of Hamiltonian systems \cite{arn1,arn2,kol,mos}.  The other was the theory of stochasticity of dynamical systems (loosely called ``chaos theory''), which demonstrated the stability of strongly irregular dynamics under small perturbations \cite{anosov,sin,smale}.

Typical Hamiltonian systems have mixed dynamics, with islands of stability (``KAM islands'') situated in a chaotic sea.  However, rigorous mathematical investigations of such systems are notoriously difficult because different analytical methods have been developed for systems with fully regular or fully chaotic dynamics.  Both approaches fail at the boundaries between chaotic and regular regions.  Numerical investigations of systems with mixed dynamics are also difficult for the same reason (small islands are not easy to find/observe numerically).  

There have been numerous attempts to find Hamiltonian systems with mixed dynamics (divided phase space) that allow an exact, rigorous analysis \cite{oval,saito,stephan}.  Recently, unexpectedly simple and visual examples of such systems were found in the form of mushroom billiards, whose geometry generalizes the long-studied stadium billiard \cite{mush,mush2}.  From a mathematical perspective, the discovery of mushroom billiards has now made it possible to address some delicate questions about the dynamics of systems with coexisting KAM islands and chaotic regions \cite{mush,mush2,stick}.  Such theoretical studies can be readily tested experimentally in many physical situations.  For example, two-dimensional billiards in essentially arbitrary geometries corresponding to systems with integrable, chaotic, and mixed dynamics can be constructed using microwave cavities \cite{stock1,stock2,sridhar}, quantum dots \cite{marcus}, and atom optics \cite{atomoptics,kaplan,raizen}.  The problems that can be studied by experimentalists using billiard systems are both fundamental and diverse, ranging from the decay of quantum correlations \cite{kaplan} to investigations of the dynamics of Bogoliubov waves for Bose-Einstein condensates confined in various billiard geometries \cite{raizen}.

In this work, we investigate chaos-chaos transitions in classical billiard systems.  To do this for a given billiard, we gradually perturb its geometry, just as one would for the order-chaos transitions that have long been studied in Hamiltonian systems.  In this case, however, the billiard's phase space has chaotic regions both before and after the perturbation rather than just after it.  We examine the proliferation and destruction of KAM islands that result from these perturbations and consider the concomitant disappearance and appearance of chaotic regions in phase space.  In particular, we investigate the example of one-particle generalized mushroom billiards.  We also study few-particle billiards, in which the confined particles collide elastically not only against the boundary but also against each other.  Towards this end, we study them in geometries whose associated one-particle billiard dynamics are integrable, chaotic, and mixed.  We consider particles of different sizes to examine the limit in which the particles become smaller and the few-particle system becomes more like the associated one-particle billiard.

Traditional mathematical studies of systems of particles deal either with just one particle (billiards and perturbations/modifications thereof) or with an infinite number of particles.  Numerical studies have been concerned with hundreds or thousands (or more) interacting particles \cite{dellago,alder,posch}.  Another tradition for studies of systems of interacting particles is to focus on the potential governing interactions between particles rather than on the shape of the containers in which the particles are confined; systems of interacting particles have traditionally been considered in a box or a torus.

However, from both the theoretical and practical standpoints, systems of just a few particles are getting increasing attention, in large part because of nanoscience and possible applications in nanotechnology.  From an experimental perspective, collisions between particles can be studied using the framework of cold atoms \cite{atomoptics,cornell}  On the theoretical side, it was recently demonstrated that stationary distributions of systems of noninteracting particles in containers with nontrivial shapes are typically very inhomogeneous \cite{mush2}.  A natural question that consequently arises is whether such phenomena may occur in systems of interacting particles.  In this paper, we give a positive answer to this question by demonstrating that the same effect is present in systems of two, three, and four hard balls in a circle.

The structure of this paper is as follows.  First, we examine the dynamics of one-particle billiards.  We give some background information on integrable and chaotic billiards before turning our attention to billiard systems with mixed regular-chaotic dynamics.  We consider, in particular, billiard geometries shaped like mushrooms and generalized mushrooms.  We then study the dynamics of two-particle billiards in containers whose corresponding one-particle billiard dynamics are integrable, chaotic, and mixed.  We subsequently examine three-particle and four-particle billiards.

\section{One-Particle Billiards}

Classical (one-particle) billiard systems are among the best-studied Hamiltonian dynamical systems \cite{sinai,predrag}.  The dynamics of a billiard are generated by the free motion of a particle (usually taken to be a point) inside a closed domain $Q$ with piecewise smooth boundary $\partial Q$ in Euclidean space $\mathbb{R}^d$.  (Most studies concentrate on the case $d = 2$.)  The confined particle collides elastically against the boundary, so the angle of incidence equals the angle of reflection.

The billiard flow $S^t$ has phase space $\mathcal{M} = \{(q,v):q \in Q, \|v\| = 1\}$.  The unit normal vector (pointing towards the interior of the billiard) to $\partial Q$ at the regular point $q \in \partial Q$ is denoted $n(q)$.  Also, let $z := (q,v)$ denote a point in phase space.

A natural projection of $\mathcal{M}$ into its boundary is $M = \{(q,v): q \in \partial Q, \|v\| = 1, \langle v,n(q)\rangle \geq 0\}$, where $\langle \cdot , \cdot \rangle$ denotes the standard inner product.  Define the billiard map $T$ by $T(q,v) = (q_1,v_1)$, where $q_1$ is the point of $\partial Q$ at which the oriented
line through $(q,v)$ first hits $\partial Q$ and $v_1 = v - 2\langle n(q_1),v\rangle n(q_1)$ is the velocity vector after the reflection against the boundary at the point $q_1$. (The billiard map is not defined if $q_1$ is at a singular point of $\partial Q$.)  When $d = 2$, this map is parametrized according to the arclength $s$ along $\partial Q$ (measured from an arbitrary point on $\partial Q$) and the angle $\varphi$ between $v$ and $n(q)$.  Hence, $\varphi \in [-\pi/2,\pi/2]$ and $\langle n(q),v\rangle =
\cos \varphi$.  The billiard flow conserves the volume measure $d\nu$ in $\mathcal{M}$.  The corresponding invariant measure for $T$ is $d\mu = A\cos \varphi ds d\varphi$, where $A = 1/[2 \,\mbox{length}(\partial Q)]$ is a normalization constant.

\subsection{Billiards with Fully Integrable or Chaotic Dynamics}

A curve $\Gamma \subset Q$ is called a caustic of the billiard if whenever any link of some trajectory is tangent to $\Gamma$, then all other links of the same trajectory are also tangent to $\Gamma$.  It is well-known that the configuration space of a billiard in a circle is continuously foliated by circles concentric to $\partial Q$.  A circular billiard is thus completely integrable. Elliptical billiards are similarly integrable, although they have two continuous families of caustics instead of just one.  One family is formed by trajectories tangent to confocal ellipses, and the other is formed by trajectories tangent to confocal hyperbolae.  In the limit of zero eccentricity, the ``hyperbolic" caustics disappear and the elliptical caustics become concentric circular caustics.  Semicircles and semiellipses are likewise integrable \cite{katok}.

Examples of chaotic billiards include {\it dispersing} billiards such as the Sinai and diamond containers \cite{elastic} and billiards with {\it focusing} boundaries such as stadia.  Neighboring parallel orbits diverge when they collide with dispersing components of a billiard's boundary.  In chaotic focusing
billiards, on the other hand, neighboring parallel orbits converge at first, but divergence prevails over convergence on average.  (Hence, billiards with focusing boundaries may be viewed as occupying an intermediate position between dispersing and integrable billiards.)  Divergence and convergence are balanced in integrable billiards.

The investigation of chaotic billiards has a long history.  The flow in classical chaotic billiards is hyperbolic, ergodic, mixing, and Bernoulli \cite{elastic,sinai}.  Autocorrelation functions of quantities such as particle position and velocity of confined particles decay exponentially \cite{garrido,posch}.  Additionally, the quantizations of classically chaotic billiards bear the signatures of classical chaos in, for example, the distributions of their energy levels, the ``scarring''/``antiscarring" (increased/decreased wavefunction density $|\psi|^2$) that corresponds to classical unstable/stable periodic orbits, and the structure of nodal curves (at which the density vanishes) \cite{gutz,oldmac,mac}.

\subsection{Billiards with Mixed Dynamics}

Typical Hamiltonian systems exhibit mixed regular-chaotic dynamics (divided phase space).  Examples of billiard geometries that yield such dynamics include non-concentric annuli (with circular boundaries)
\cite{saito} and ovals \cite{oval}.  For systems with mixed dynamics, however, it is difficult to exactly determine and describe the structure of the boundaries of stability islands.  A container shaped like a mushroom provides an example of a billiard with divided phase space for which such precise mathematical analysis is feasible \cite{mush,mush2,stick}.  One reason it is natural to investigate mushroom billiards and their generalizations is that they can be designed in a precise manner so that they have an arbitrary number of integrable and chaotic components, each of which occupies the desired fraction of the phase space volume.  Their study may thus lead to a better understanding of Hamiltonian systems with divided phase space.

A circular mushroom billiard consists of a semicircular cap with a stem of some shape attached to the cap's base.  Circular mushroom billiards provide a continuous transition between integrable (semi)circular billiards and chaotic (semi)stadium billiards.  In circular mushrooms, trajectories that remain in the cap are integrable, whereas those that enter the stem are chaotic (except for a set of measure zero).  The precise characterization of all such trajectories is completely understood, and one can change the dimensions of the mushroom to controllably alter the relative volume fractions in phase space of initial conditions leading to integrable and chaotic trajectories.  Another interesting property of circular mushroom billiards, which was demonstrated recently, is that they exhibit ``stickiness'' in their chaotic trajectories (characterized by long tails in recurrence-time statistics) even though they do not possess hierarchies of KAM islands \cite{stick}.

\begin{figure}
                \centerline{
                \includegraphics[width=0.4\textwidth]{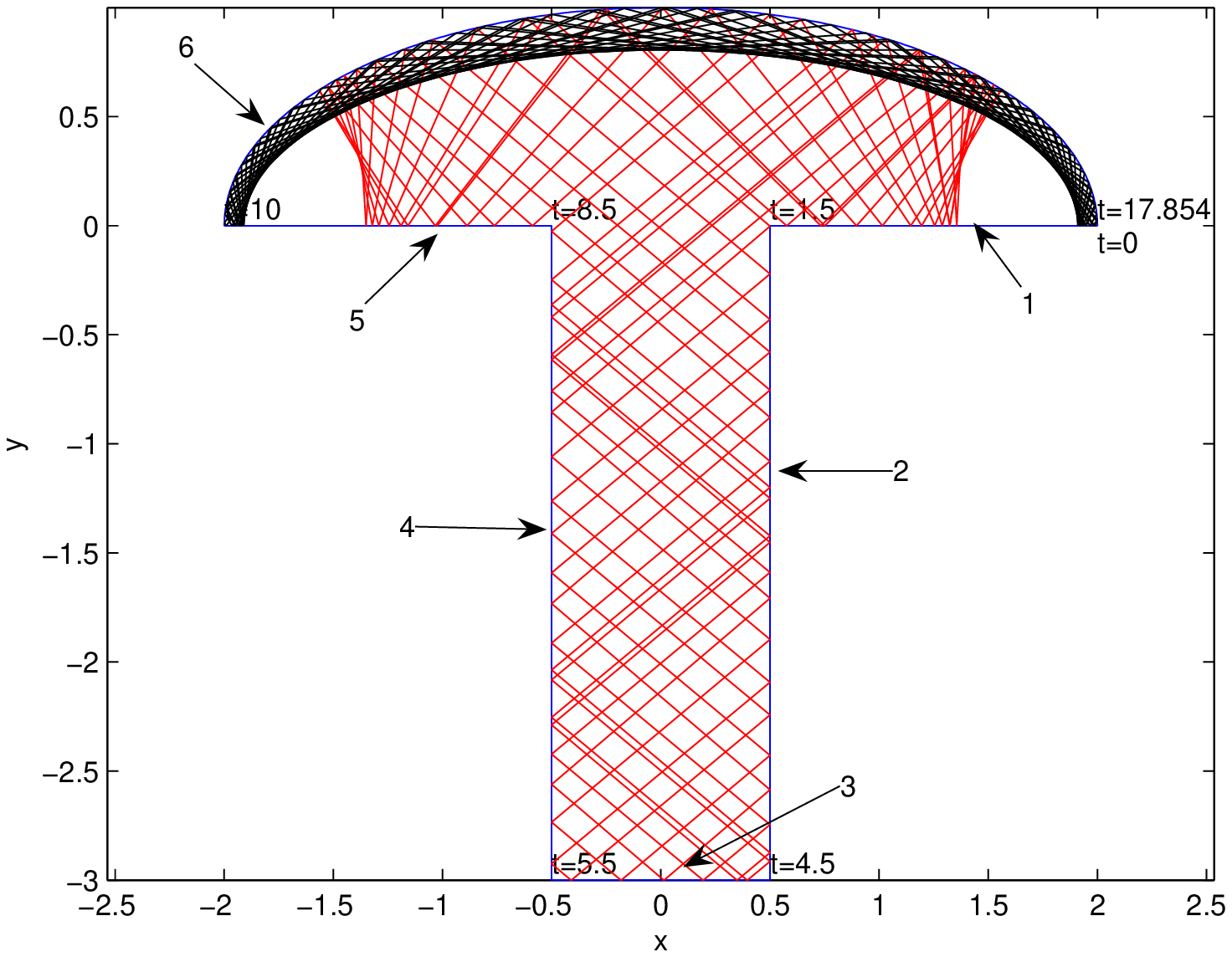}
                \hspace{.2 cm}
                \includegraphics[width=0.4\textwidth]{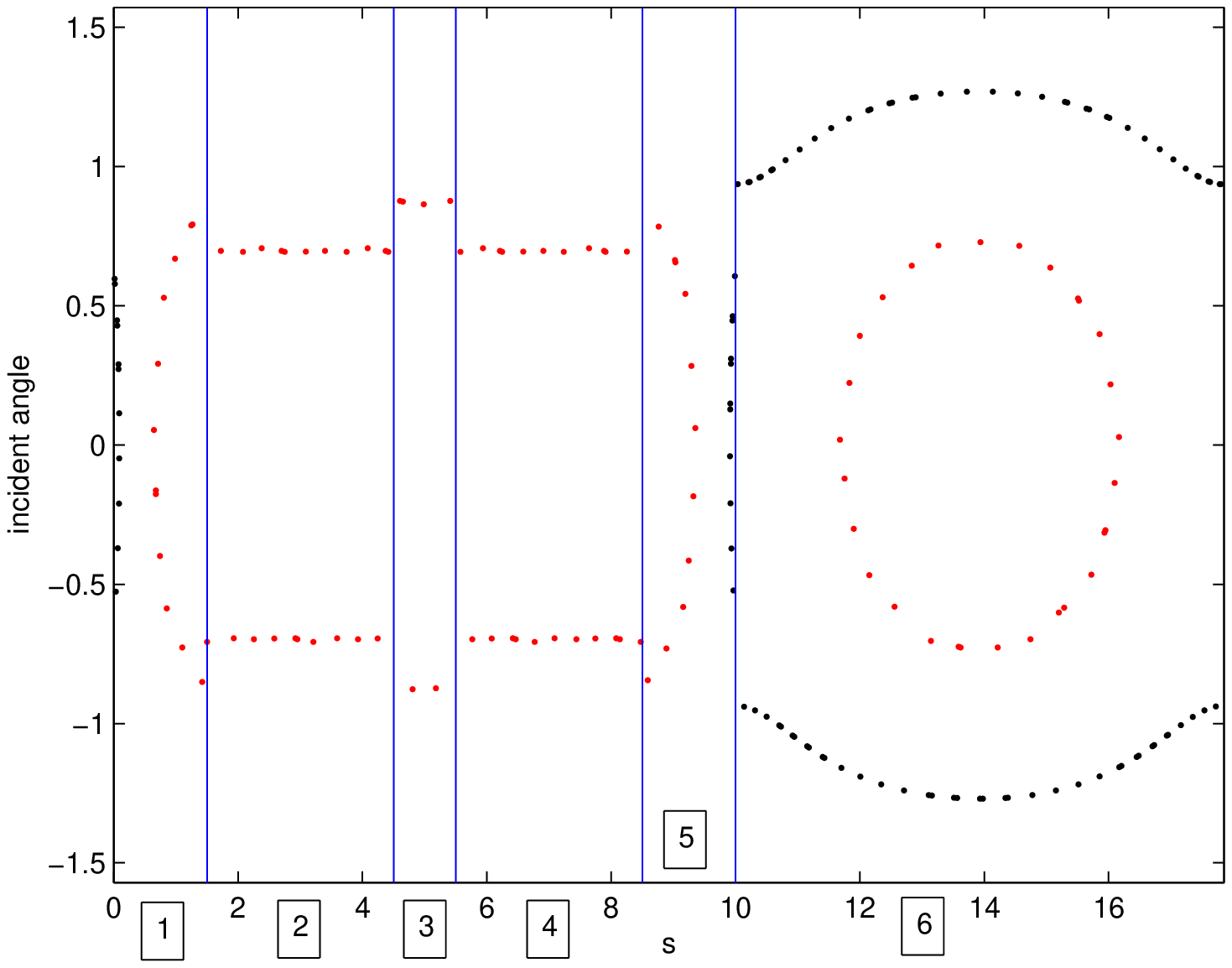}
                                }


                \caption{(Color online) Configuration space (a) [in continuous time] and phase space (b) [in discrete time] plots for an elliptical mushroom billiard with a rectangular stem. The semimajor and semiminor axes of the mushroom's elliptical cap have respective lengths of $2$ and $1$.  The stem has height $3$ and width $1$.  In panel (b), the horizontal axis indicates the location of a boundary collision, as measured by the arclength in the clockwise direction from the right-most point of the table.  The vertical axis indicates the incident angle of the particle's collision with the boundary.  To illustrate how to read subsequent figures in this paper, we have plotted a trajectory (in black) that enters the mushroom's stem and another one (in red/gray) that does not.  We label the regions in (b) to help identify their corresponding boundary arcs in (a).  (Regions matched to different arcs are separated in (b) by vertical lines, which correspond to singular points of the billiard boundary.)  For example, collisions against the elliptical arc of the mushroom's cap occur in region 6 (the right-most region) of phase space.}
\label{demo}
\end{figure}

In this section, we investigate (axially symmetric) elliptical mushroom billiards, which generalize the elliptical stadium billiard \cite{bern,lopac} and consist of a semiellipse with a stem extending from the center of its major (horizontal) axis.  First, we show with an example plot (Fig.~\ref{demo}) how to visualize the dynamics in both configuration space and phase space.  Utilizing mushrooms with several stem geometries (rectangular, triangular, and trapezoidal), we subsequently examine symmetric and asymmetric periodic orbits and illustrate the birth and death of KAM islands that occur as the stem height is increased.  As an example, we derive analytically the stem height at which the period-2 orbit along the mushroom's axis of symmetry becomes unstable.

\begin{figure}
                \centerline{
                (a)
                \includegraphics[width=0.4\textwidth]{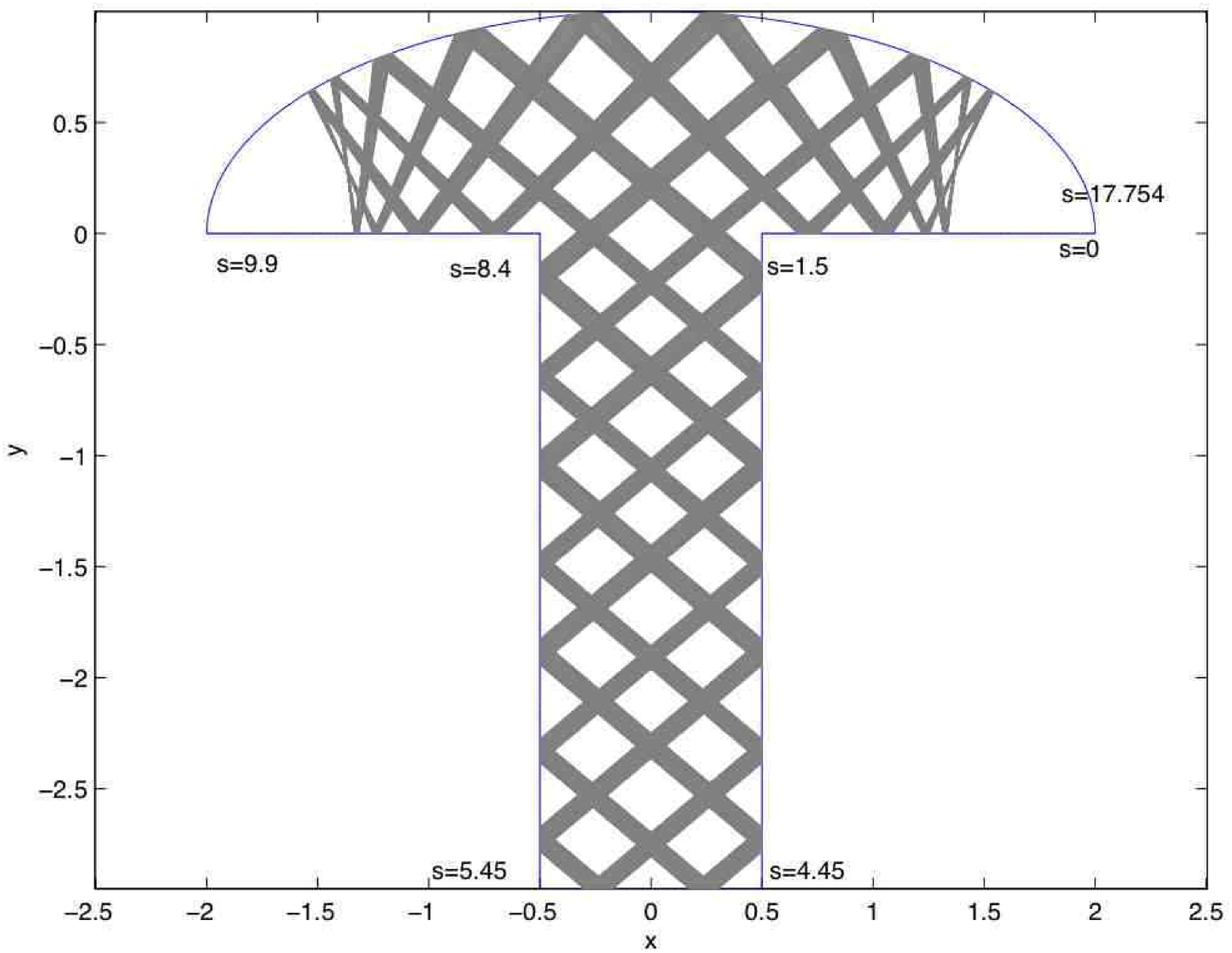}
    \hspace{.2 cm}
                (b)
                \includegraphics[width=0.4\textwidth]{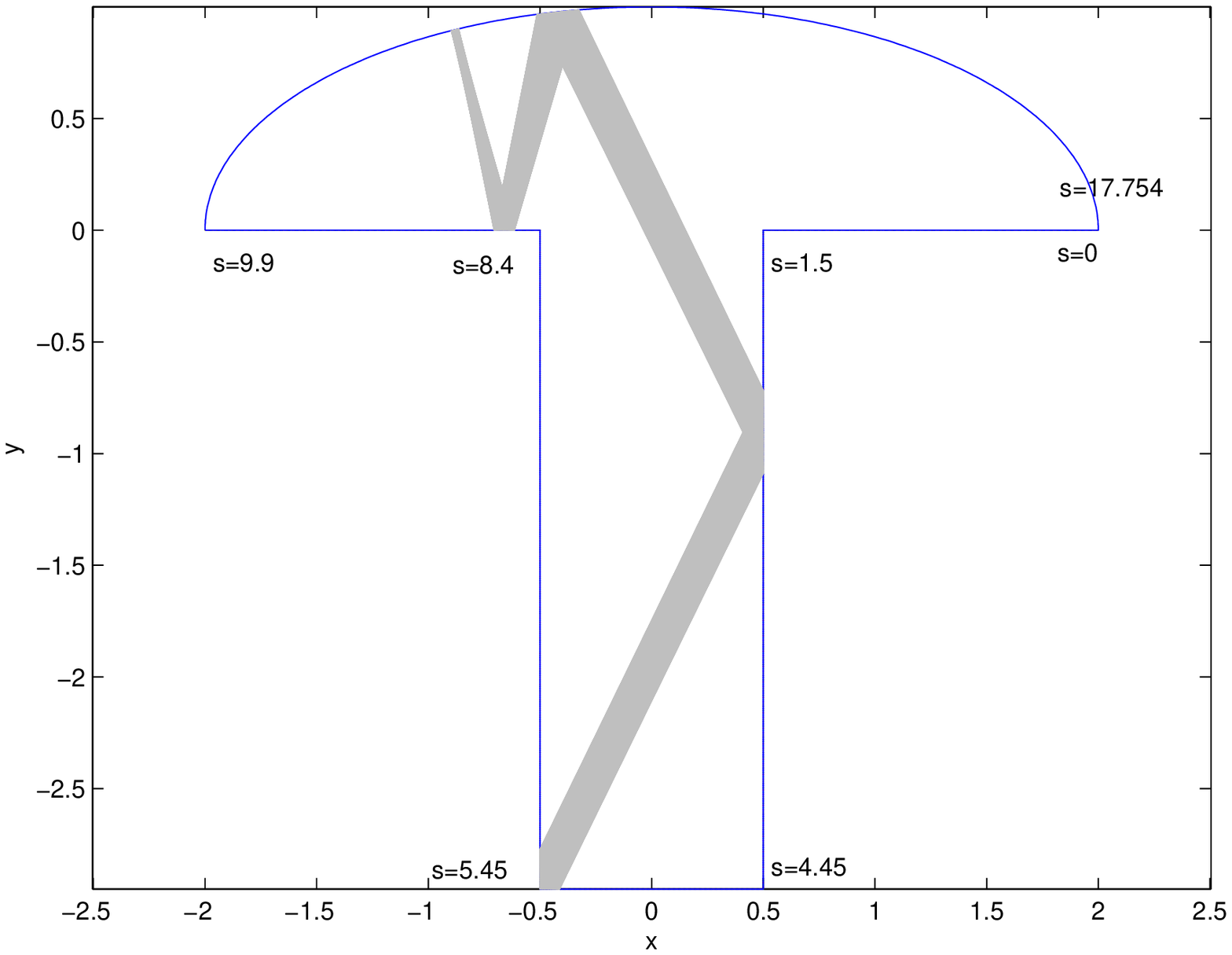}}


                \caption{Periodic orbits of the elliptical mushroom from Fig.~\ref{demo}. (a) Trajectory near a stable, symmetric periodic orbit of period $34$ (medium gray).  (b) Trajectory near a stable, asymmetric periodic orbit of period $9$ (light gray).}
\label{ellipse2}
\end{figure}

To study the dynamics of elliptical mushroom billiards, we conducted extensive numerical experiments using software that we wrote and have made publicly available \cite{billiardsim}.  To illustrate subsequent plots, we depict in Fig.~\ref{demo} the billiard system in configuration and phase space for the case of a rectangular stem.  We then show the dynamics of two of this billiard's trajectories in Fig.~\ref{ellipse2}.  The phase space of this billiard is divided, as it has both regular and chaotic regions.  In mushroom billiards with rectangular stems, we observe stable periodic orbits of arbitrary length.

\begin{figure}
                \centerline{
                (a)
                \includegraphics[width=0.4\textwidth]{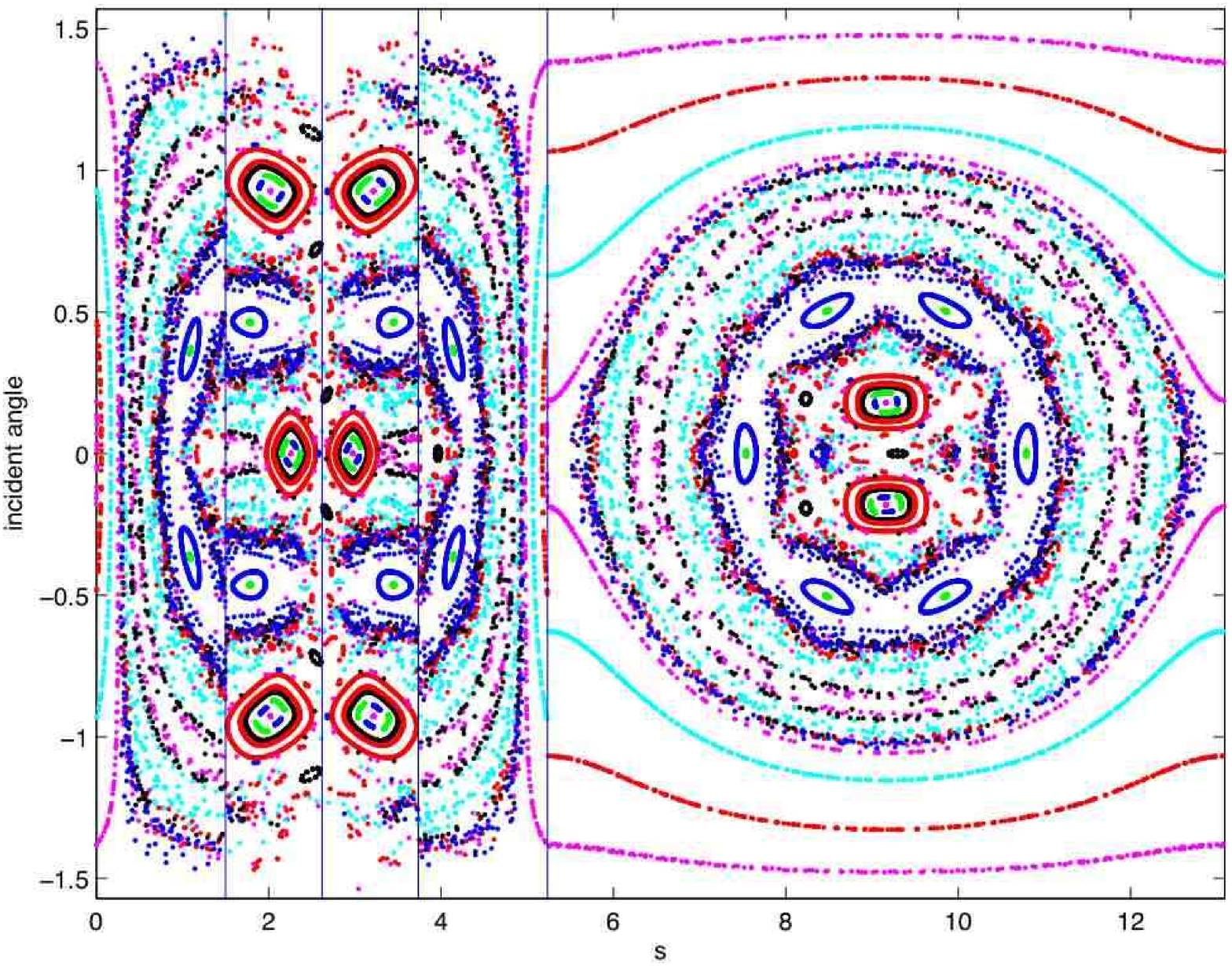}
                \hspace{.2 cm}
                (b)
                \includegraphics[width=0.4\textwidth]{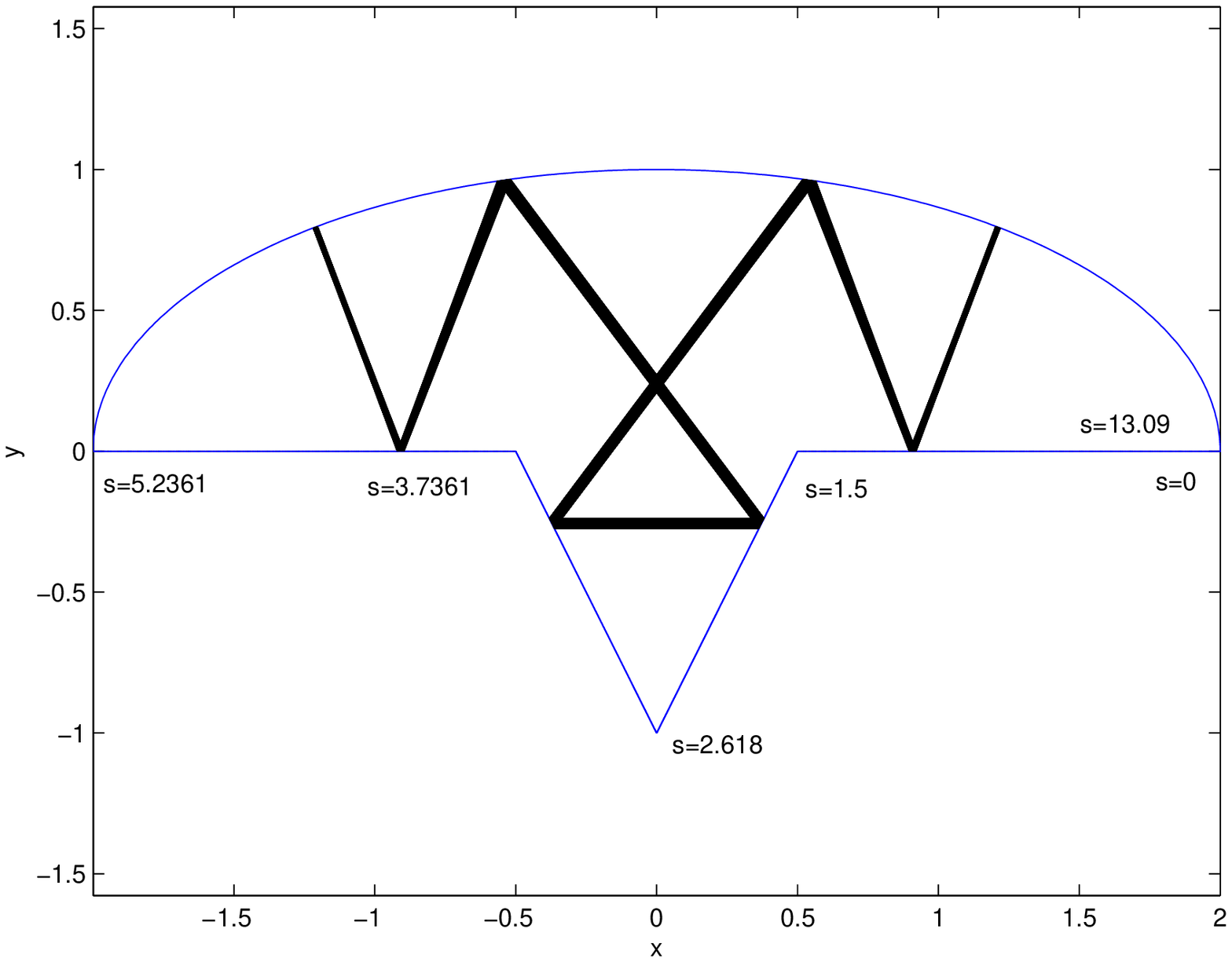}}
                \centerline{
                (c)
                \includegraphics[width=0.4\textwidth]{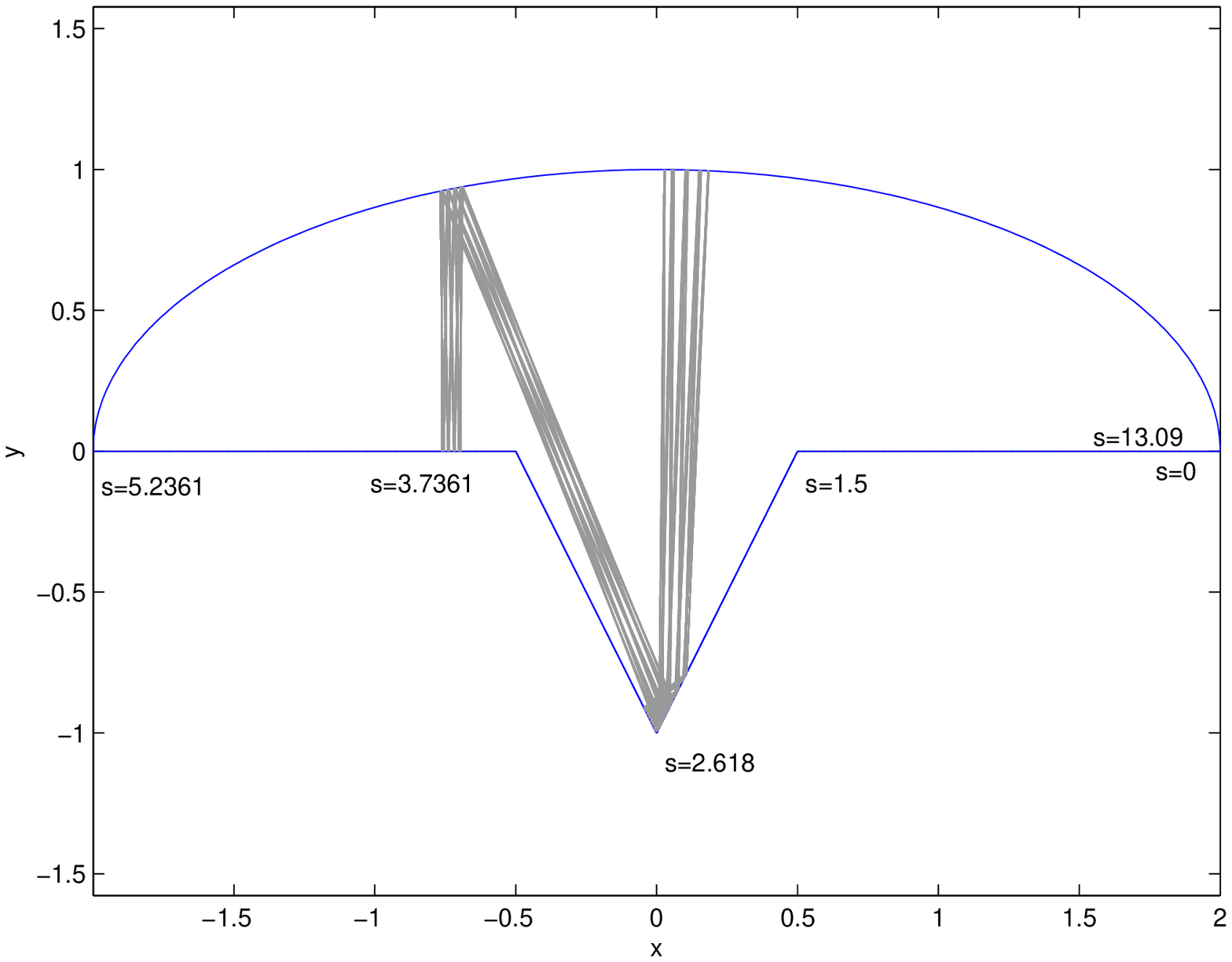}
\hspace{.2 cm}
                (d)
                \includegraphics[width=0.4\textwidth]{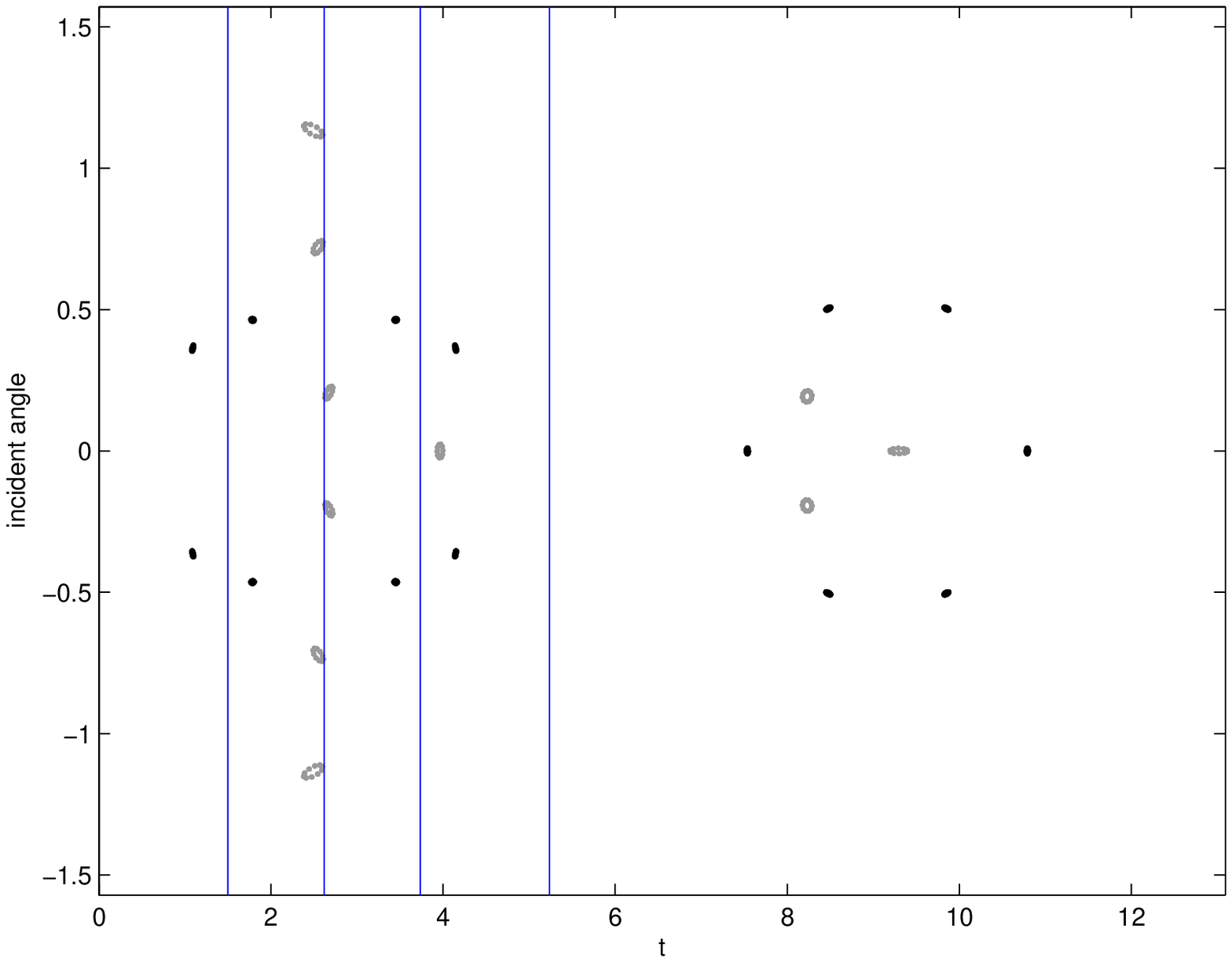}}


                \caption{(Color online) Divided phase space and periodic orbits of an elliptical mushroom billiard with a stem shaped like an isosceles triangle.  The semimajor and semiminor axes have lengths 2 and 1, respectively.  The triangular stem has a height and width of 1.  (a) Phase space showing multiple islands.  The figure depicts 13 initial conditions.  (b) Trajectory near a stable, symmetric periodic orbit of period 14 (shown in black).  (c) Trajectory near a stable, asymmetric periodic orbit of period 10 (shown in gray).  (d) Phase space depicting the periodic orbits from (b) and (c).}
\label{triangle}
\end{figure}

We also study elliptical mushrooms with triangular (see Fig.~\ref{triangle}) and trapezoidal stems (no figures shown).  When outside the ``extended stem" of the mushroom, trajectories behave in the same manner regardless of the shape of the stem.  (The extended stem of a mushroom billiard consists of the stem itself plus the portion of the hat inside the largest geometrically similar semiellipse concentric to the hat that touches an upper corner of the stem.)  Differences in dynamics for mushrooms with different stems manifest only for trajectories that enter the extended stem (and hence the stem per se).  We remark that mushroom billiards with trapezoidal stems possess nontrivial periodic orbits that never leave the stem.  In contrast, there are no such periodic orbits for mushroom billiards with triangular stems and only trivial orbits of this kind for mushrooms with rectangular stems.

Symmetric mushroom billiards with rectangular, triangular, and trapezoidal stems all have stable symmetric and asymmetric periodic orbits that traverse both the hat and the stem.  For triangular and trapezoidal stems, asymmetric periodic orbits that enter the stem are one of two types: (1) the trajectory collides against the stem near one of its bottom corners; or (2) the trajectory collides against a line segment that is perpendicular to the current link of the trajectory.  (In the latter case, the particle then turns around and begins to trace its path backwards.)  Rectangular stems, however, only have orbits of type 1.  These orbits return to the cap from the stem at the same angle (with a vertical reflection) at which they left, with resulting boundary collisions in the cap as if the trajectory had bounced once off a linear segment perpendicular to the trajectory rather than bouncing several times in the stem (see, for example, Fig.~\ref{ellipse2}b).  Such trajectories enter the stem, bounce around against the stem's boundary, and then emerge from it in the same fashion that they entered.  By contrast, type 1 asymmetric periodic orbits in triangular stems do not have this property (see, for example, Fig.~\ref{triangle}c).  This difference in dynamics reveals the presence of periodic orbits in mushroom billiards with rectangular stems that are not present in ones with triangular stems.  For symmetric periodic orbits, we observe a similar reflection property for all three types of stems, except that there is an additional horizontal reflection in the trajectory's return from the stem to the cap.  Such trajectories (see Figs.~\ref{ellipse2}a and \ref{triangle}b) result in periodic orbits that are present in all three families of mushroom billiards.  No matter what they do inside the stem, such orbits leave the stem with the same angle (ignoring sign) with respect to the mushroom's axis of symmetry that they had when they entered.

Numerous bifurcations occur with changes in stem height, as KAM islands rapidly proliferate and change configurations.  For instance, using triangular stems of width $1$ and heights ranging from
$0.1$ to $2.0$ in increments of $0.1$, we observed a large variation in the number of central islands appearing vertically down the left side of phase space (corresponding to collisions against the billiard stem).  For example, there are 6 such islands in Fig.~\ref{triangle}a.  In some configurations, the islands correspond to two different trajectories: an asymmetric trajectory and its horizontal reflection.  In other configurations, one symmetric trajectory hits all the islands.  We observed similar phenomena for trapezoidal stems with lower width $1.5$, upper width $0.5$, and heights ranging from $0.1$ to $2.0$ (in increments of $0.1$).

\begin{figure}
                \centerline{
                (a)
                \includegraphics[width=0.4\textwidth]{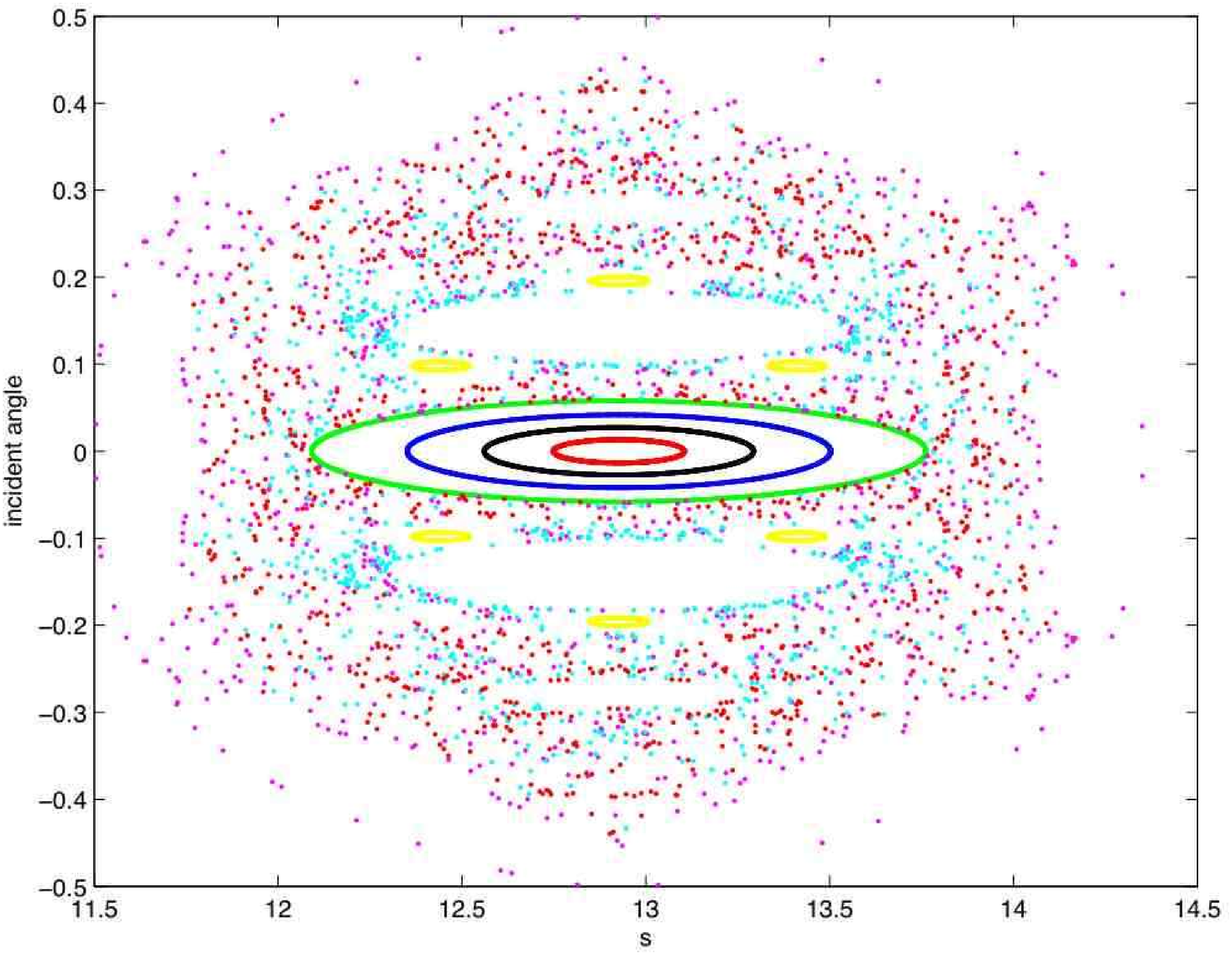}
\hspace{.2 cm}
                (b)
                \includegraphics[width=0.4\textwidth]{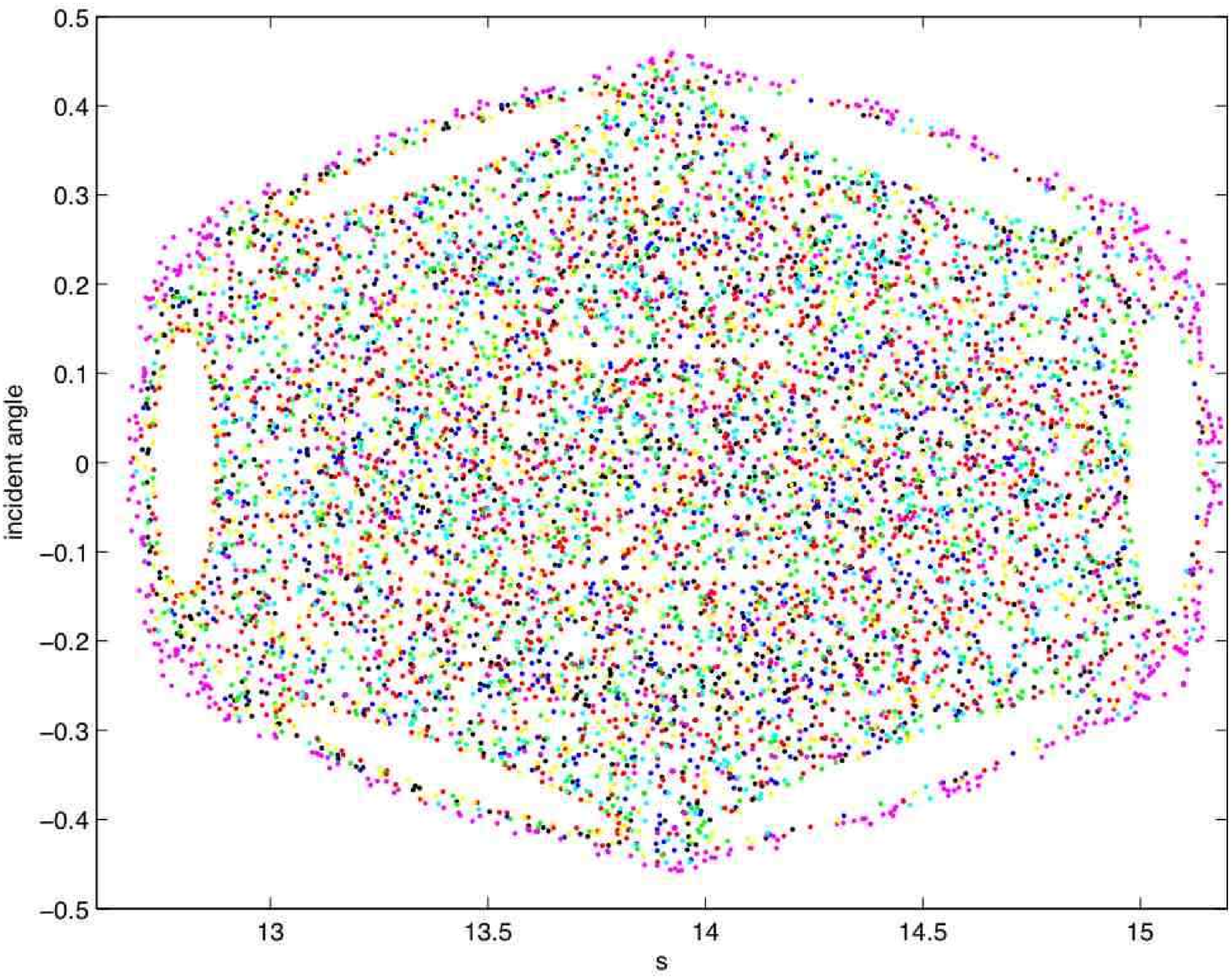}}


                \caption{(Color online) Magnifications of phase space showing the disappearance of the central KAM island---which corresponds to the vertical period-2 trajectory passing through the axis of symmetry of the elliptical mushroom---as a result of varying the stem height.  The example depicted has a semimajor axis of length 2, a semiminor axis of length 1, and a rectangular stem of width 1.  The central island disappears at stem height 3 (when the trajectory becomes unstable).  The figure panels show an enlargement of the portion of phase space corresponding to collisions with the elliptical arc.  (a) Stem height 2.5.  (b) Stem height 3.}\label{stability}

\end{figure}

The bifurcation from stability to instability for the periodic vertical trajectory in symmetric elliptical mushrooms (for stems with a neutral arc at the bottom) occurs when the distance between the bottom of the stem and top of the cap is twice the radius of curvature of the cap.  (More generally, this result applies to period-2 orbits which collide alternatively with a focusing elliptical arc and a straight (neutral) arc, which also occurs, for example, in the so-called "flattened" elliptical semistadium billiards with two foci \cite{lopac}.)  The eigenvalues of the stability matrix (Jacobian) for this period-2 orbit are
\begin{equation}
    \lambda_{1,2} = -\frac{1}{2} + \frac{z}{2} \pm \frac{1}{2}\sqrt{-3 - 2z + z^2}\,,
\end{equation}
which are complex conjugates when the stem height $z < 3$ and real (with one eigenvalue having magnitude greater than unity) when $z > 3$.  This vertical trajectory thus becomes unstable at $z = 3$.  This result is illustrated numerically in Fig.~\ref{stability}, which depicts an enlarged portion of phase space corresponding to collisions with the elliptical arc.  The central island corresponding to the vertical trajectory shrinks as the stem height is increased until it finally disappears at height $3$.


Bunimovich proved that elliptical mushroom billiards with sufficiently long stems have one chaotic component and two integrable islands provided (1) the stem does not intersect the edges of the cap and (2) the stem does not contain the center of the cap's base \cite{mush}.  (If the stem is not sufficiently long, other islands can also appear.  One can thus observe such islands, for example, in elliptical stadium billiards \cite{lopac}.)  If condition (1) does not hold, the billiard contains one integrable island, formed by trajectories in the cap tangent to hyperbolae.  If condition (2) does not hold, the elliptical mushroom contains only the island formed by trajectories in the cap tangent to ellipses.  If both conditions are violated, then no integrable islands exist.  In the limit of zero eccentricity, one obtains a circular mushroom billiard and the islands resulting from condition (2) disappear forever.  Consequently, circular mushrooms contain no integrable islands if the stem intersects the edge of the cap.

\section{Two-Particle Billiards}

We consider here the dynamics of two identical hard balls confined in containers whose one-particle billiard counterparts have integrable, chaotic, and mixed dynamics.  In these systems, the confined particles collide elastically not only against the boundary but also against each other.  We studied particles of various sizes to examine the ``billiard" (low density) limit in which the particle radius vanishes.  We find that although the two-particle dynamics are chaotic for all three situations, two-particle billiards with a chaotic or mixed billiard limit have different dynamics from those with an integrable billiard limit.

The configuration space of a hard ball of radius $r$ in a container $Q \subset \mathbb{R}^d$ is equivalent to that of a point particle in a smaller container $Q_0 \subset Q$.  The boundary $\partial Q_0$ is formed by all the points $q \in {\rm int}(Q_0)$ located a distance $\rho$ from the boundary $\partial Q$.  In the case of $N$ noninteracting hard balls (that pass through each other and collide only against the boundary), the resulting configuration space is the direct product $Q_N := Q_0 \times \cdots \times Q_0$ ($N$ times) and the phase space similarly consists of $N$ copies of the one-particle phase space.

For $N$ interacting particles, it is difficult in general to explicitly describe the configuration space $Q_N$,  which can have a very complicated topology.  For hard balls, the interaction potential is infinite if the particles collide and zero otherwise.  To obtain the configuration space, one thus removes from $Q_N$ the $N(N-1)/2$ cylinders corresponding to pairwise interactions between particles \cite{mush2}.  (One obtains these cylinders by considering collisions between two fixed particles and allowing all others to move (without colliding) inside $Q$.)

\subsection{Two Particles in Geometries Corresponding to Integrable Billiards}

First, we consider two-particle dynamics in configurations which are integrable for point-particle billiards with a single particle.  Using a confining circular boundary (of unit radius), we examined particles with radius $1/4$, $1/10$, and $1/20$ to study the dynamics as one approaches the underlying integrable billiard (i.e., as there are successively fewer collisions between the two particles).

\begin{figure}
                \centerline{
                (a)
                \includegraphics[width=0.25\textwidth]{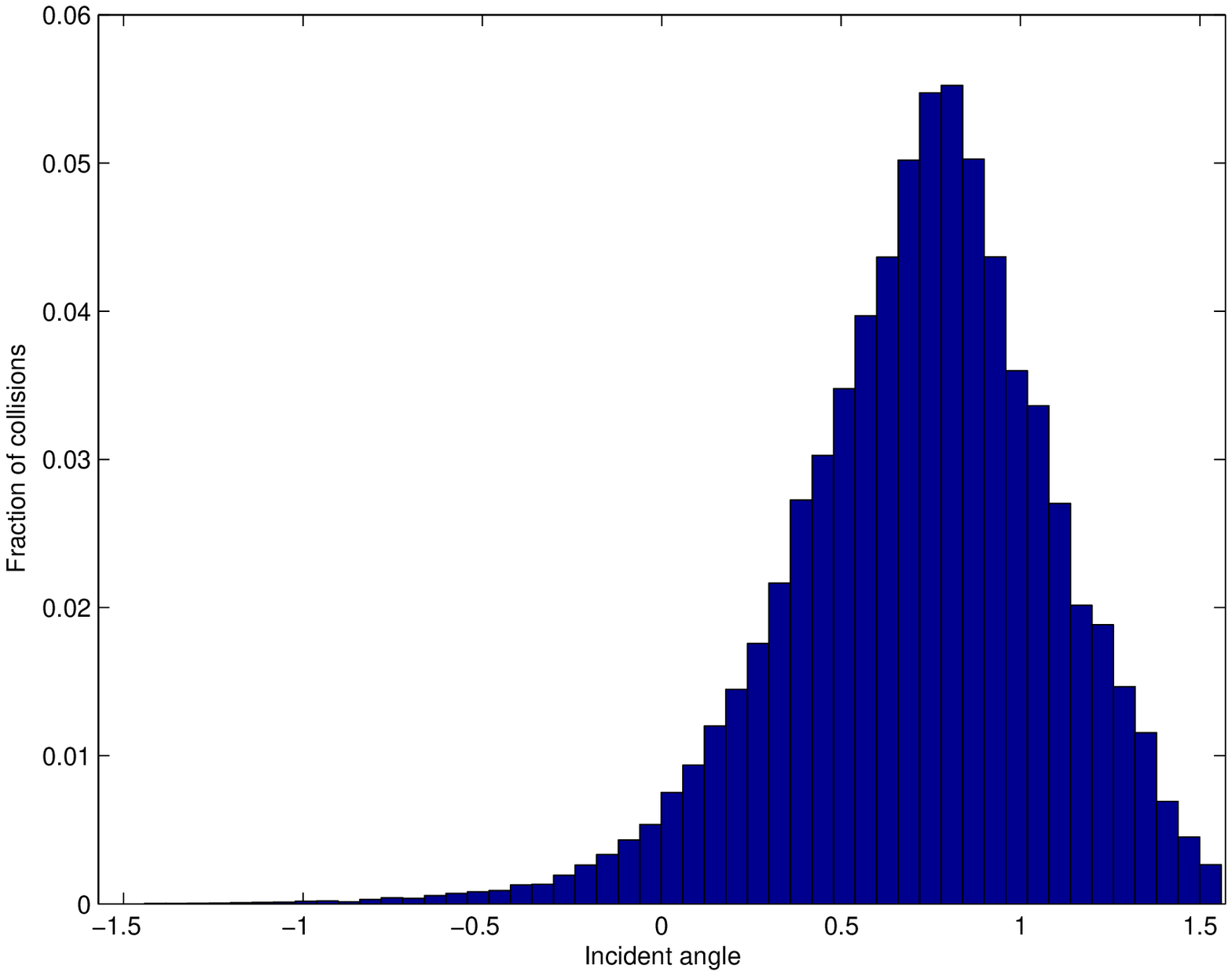}
                \hspace{.2 cm}
                (b)
                \includegraphics[width=0.25\textwidth]{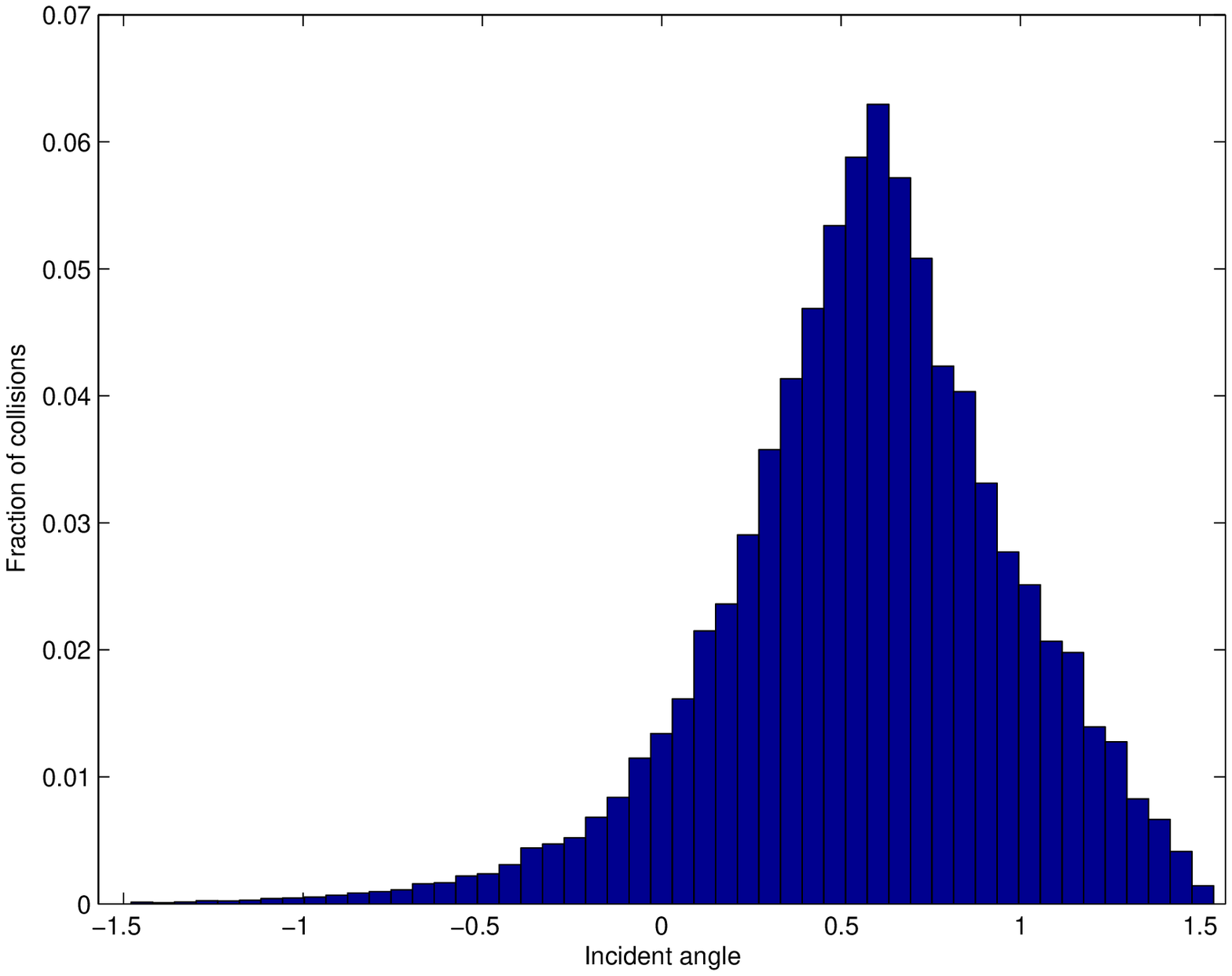}
                \hspace{.2 cm}
                (c)
                \includegraphics[width=0.25\textwidth]{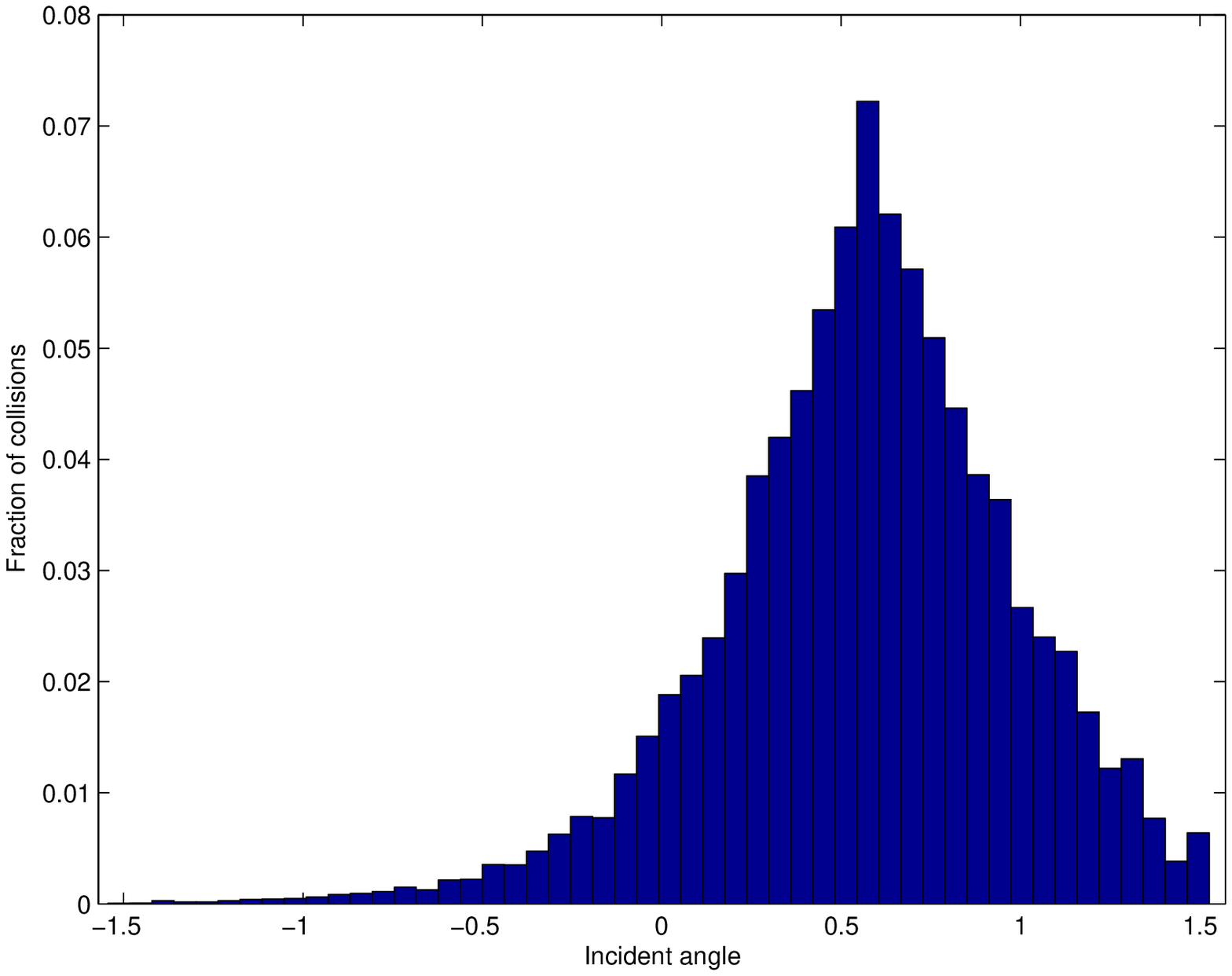}}

                \caption{Distribution of incident angle for collisions between particles and boundary for two identical particles of varying radius confined in a circle of radius 1.  The initial conditions for the three plots are shown in the last row of Table~\ref{tab1} and discussed in the text.  We simulate this system through 200,000 total collisions of either particle with the boundary.  (In this and all subsequent multiple-particle simulations, at most one data point is generated after each collision, as the incident angle remains the same if there are no particle--particle collisions.) (a) Particle radius $1/4$.  (b) Particle radius $1/10$.  (c) Particle radius $1/20$.}
\label{circle}

\end{figure}

In Fig.~\ref{circle}, we show the distribution of incident angles for boundary collisions with two identical particles confined in a circle of radius 1.  (The incident angle with the boundary is a natural observable for multiple-particle circular billiards, as it is a constant of motion for one-particle circular billiards.  Because of the system's symmetry, the coordinate along the boundary is not essential.)  We observe noticeable differences in the left tail of the distribution between the radius $1/4$ particles and the two smaller particles, as tails in particle--particle versus particle--boundary collisions get heavier as the particle size becomes smaller.  For the radius $1/4$ particles, this distribution has a standard deviation of $\sigma \approx 0.3655$  By contrast, it is about $0.4024$ for radius $1/10$ and about $0.4112$ for radius $1/20$.  We also note that the collisions have a non-zero mean angle (about $0.5$), as indicated in Table~\ref{tab1}.

\begin{figure}
                \centerline{
                (a)
                \includegraphics[width=0.25\textwidth]{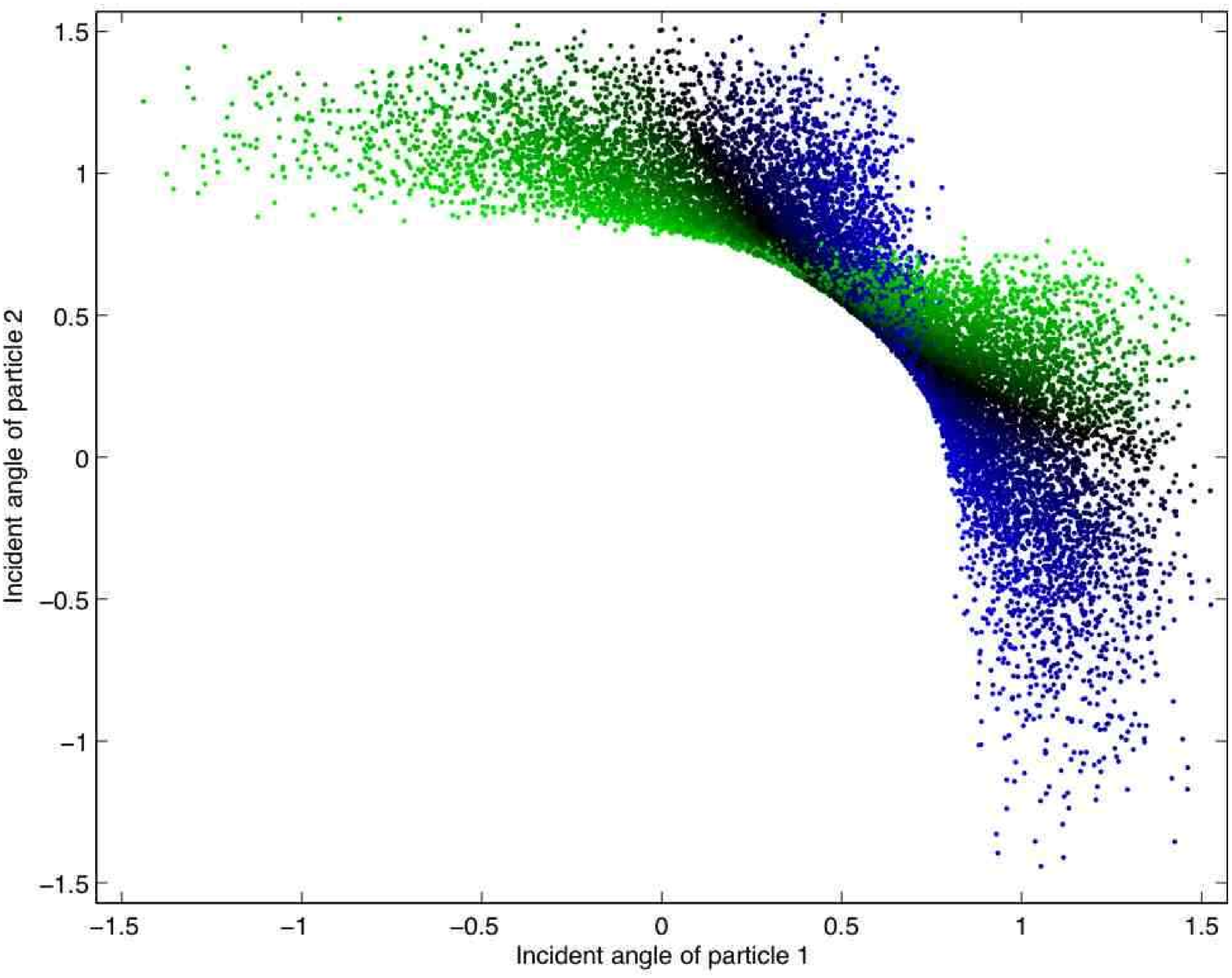}
                \hspace{.2 cm}
                (b)
                \includegraphics[width=0.25\textwidth]{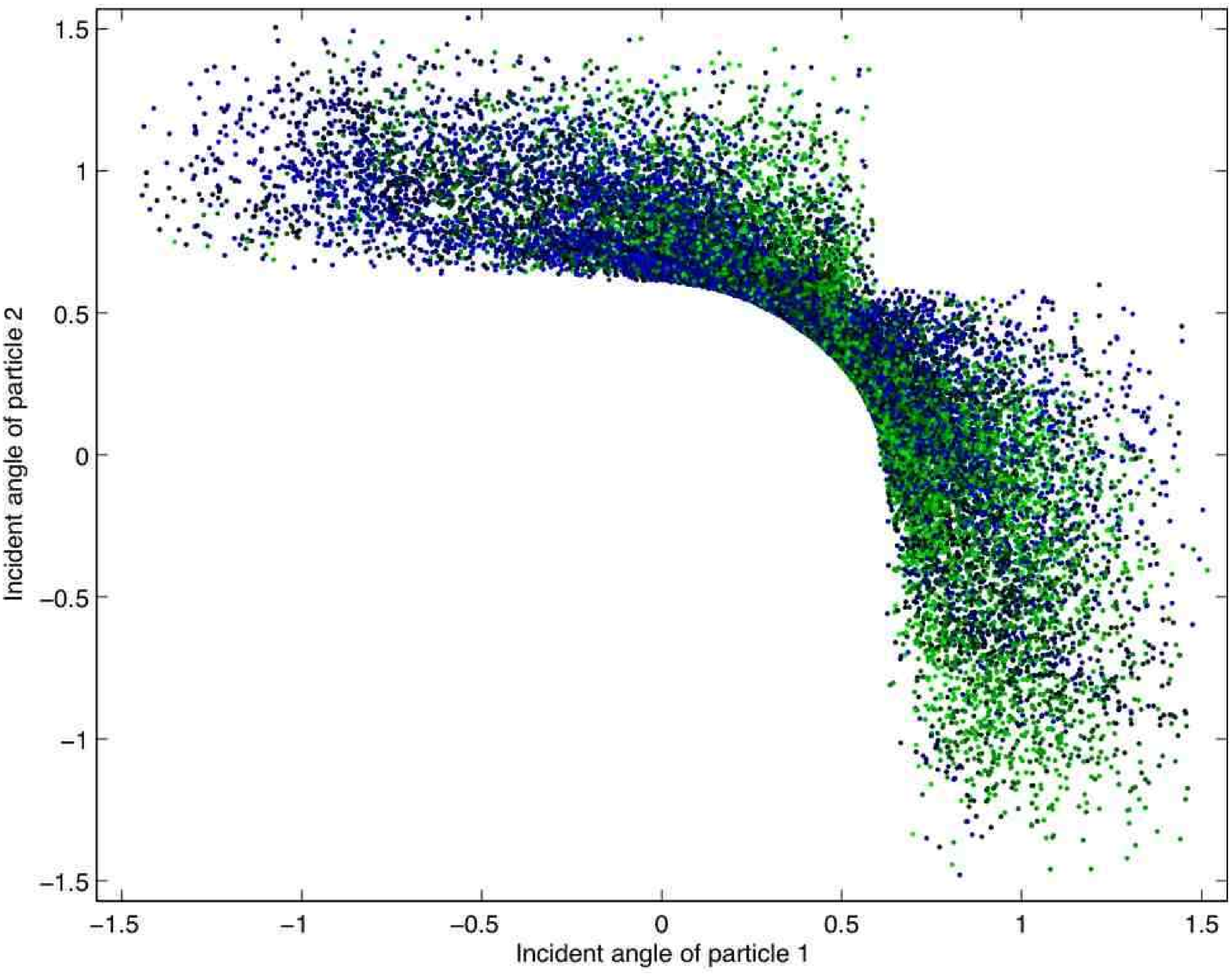}
                \hspace{.2 cm}
                (c)
                \includegraphics[width=0.25\textwidth]{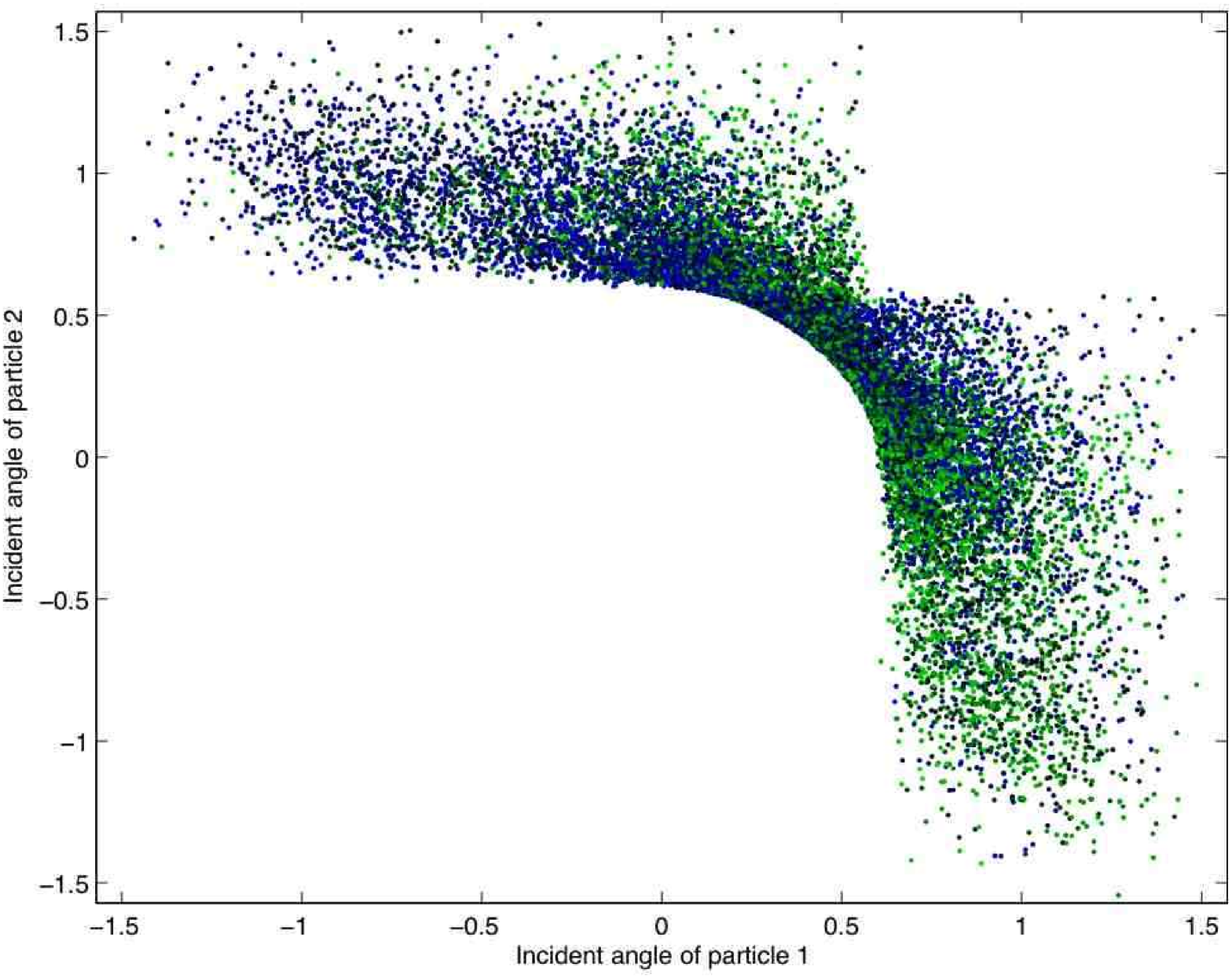}}

                \caption{Incident angle of particle 1 versus incident angle of particle 2 for data from Fig.~\ref{circle}.  The colors signify the speed ratio of particle 1 to particle 2, with blue dots indicating $v_1 > v_2$, green dots indicating $v_1 < v_2$, and black dots indicating $v_1 = v_2$.  The colors are scaled so that collisions with speed ratios close to 1 are shown as nearly black and those with larger speed disparities are brighter.  (a) Particle radius $1/4$.  (b) Particle radius $1/10$.  (c) Particle radius $1/20$.  As shown in panel (a), the particles can be tracked individually when they are sufficiently large.}
\label{special}

\end{figure}

Although we obtain the same distribution of collision angles for all three particle sizes, we show in Fig.~\ref{special} that one can separately track the two particles provided they are sufficiently large.  As the particle radii are increased relative to the container radius, their motion becomes further constrained until they can always be distinguished (tracked) even in this ergodic setting.  This is not true for general two-particle systems.  In Fig.~\ref{special}, we plot the incident angle of particle 1 versus the incident angle of particle 2 and color-code the figure based on their speed ratio $v_1/v_2$.  Bluer dots signify larger $v_1/v_2 > 1$, greener dots signify smaller $v_1/v_2 < 1$, and black dots signify $v_1/v_2 = 1$.  The colors are scaled from blue to green, so that brighter colors signify a larger speed disparity.  For example, blue-black dots indicate that particle 1 is only slightly faster than particle 2.

\begin{table}
\centerline{\begin{tabular}{|c|c|c|c|c|c|c|c|c|c|} \hline
$x_1$ & $y_1$ & $\alpha_1$ & $v_1$ & $x_2$ & $y_2$ & $\alpha_2$ & $v_2$ & $\mu$ &
$\sigma$ \\ \hline
$0.2137$ & $-0.0280$ & $-0.2735$ & $1.6428$ & $0.7826$ & $0.5242$ & $-3.0253$ &
$1.1407$ & $-0.1637$ & $0.4650$ \\
$-0.1106$ & $0.2309$ & $-0.5925$ & $1.8338$ & $0.4764$ & $-0.6475$ & $2.7361$ &
$0.7982$ & $0.2067$ & $0.4395$ \\
$0.4252$ & $-0.6533$ & $2.4586$ & $0.9129$ & $0.2137$ & $-0.0280$ & $1.6468$ &
$1.7795$ & $-0.0429$ & $0.3832$ \\
$-0.3324$ & $0.3977$ & $2.1754$ & $0.4053$ & $-0.3010$ & $-0.3720$ & $0.1580$ &
$1.9585$ & $-0.2028$ & $0.4722$ \\
$-0.7099$ & $0.0233$ & $-2.1936$ & $0.7567$ & $-0.5459$ & $-0.1875$ & $1.2434$ &
$1.8513$ & $0.1548$ & $0.4401$ \\
$0.0082$ & $0.2572$ & $0.0177$ & $0.8578$ & $-0.4308$ & $0.2477$ & $1.3161$ &
$1.8067$ & $0.5516$ & $0.5205$ \\  \hline
\end{tabular}}
\caption{Mean $\mu$ and standard deviation $\sigma$ of the incident angle for collisions between the particles and boundary for two particles of radius $1/4$ confined in a circle of radius 1.  The center of particle $i$ has initial position $(x_i,y_i)$ with initial velocity $v_i$ at angle $\alpha_i$ measured from the horizontal.  (The choice of initial conditions is discussed in the text.)}
\label{tab1}
\end{table}

We choose the initial conditions for these simulations as follows: The center of particle 1 is selected uniformly at random from all points inside the circle that are not within $1/4$ of the boundary of the circle.  The center of the second particle, which must have distance at least $1/2$ from the first initial coordinate and distance at least $1/4$ from the circle's boundary, is chosen uniformly at random from its allowed points.  We choose the angles uniformly at random between $-\pi$ and $\pi$.  We then select the first speed uniformly at random from $[0,2]$ and choose the second speed to ensure that the sum of the squares of the speeds is $4$.  

We measure the locations of collisions between two particles in terms of their distance from the center of the table.  The smaller the particle, the farther away from the origin one expects to find these collisions.  For radius $1/4$ particles, the distance from the origin peaks near $0.48$ and drops off approximately linearly on both sides of the peak until it reaches $0$ near distances $0.28$ and $0.71$.  For radius $1/10$ particles, the collision location distribution peaks at roughly $0.78$ and drops off sharply to the right; it reaches $0$ near $0.35$ and $0.9$.  For radius $1/20$ particle, collisions peak near distance $0.88$ from the center and their distribution again drops off sharply to the right, reaching $0$ near $0.38$ and $0.94$.

We investigate the distributions of the number of total particle--boundary collisions between successive inter-particle collisions, the average intervening times between which are $1.8320$, $4.8933$, and $10.3158$ for particles of radius $1/4$, $1/10$, and $1/20$, respectively.  As expected, this distribution drops off more sharply for smaller particles. For example, for radius $1/4$ particles, more than $90\%$ of inter-particle collisions occur before there are 6 consecutive collisions of a particle with the boundary.  For particles of radius $1/10$ and $1/20$, the number of consecutive collisions at this $90$th percentile are $11$ and $22$, respectively.  The modal number of boundary collisions between inter-particle collisions is $1$ in each case, with respective occurrence probabilities of $0.3361$, $0.2495$, and $0.1586$ for particles of radius $1/4$, $1/10$, and $1/20$.

\begin{figure}
                \centerline{
                (a)
                \includegraphics[width=0.2\textwidth]{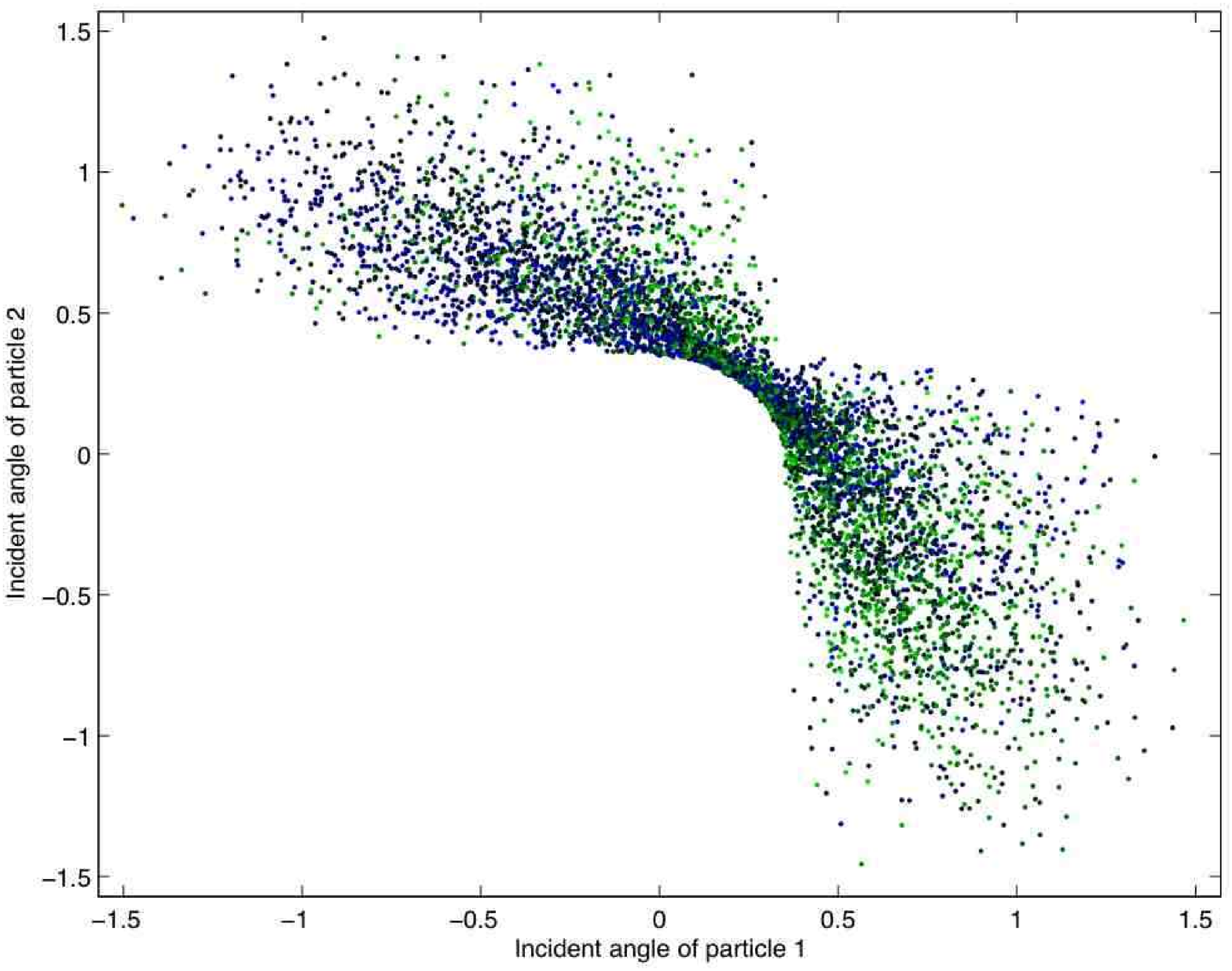}
                \hspace{.2 cm}
                (b)
                \includegraphics[width=0.2\textwidth]{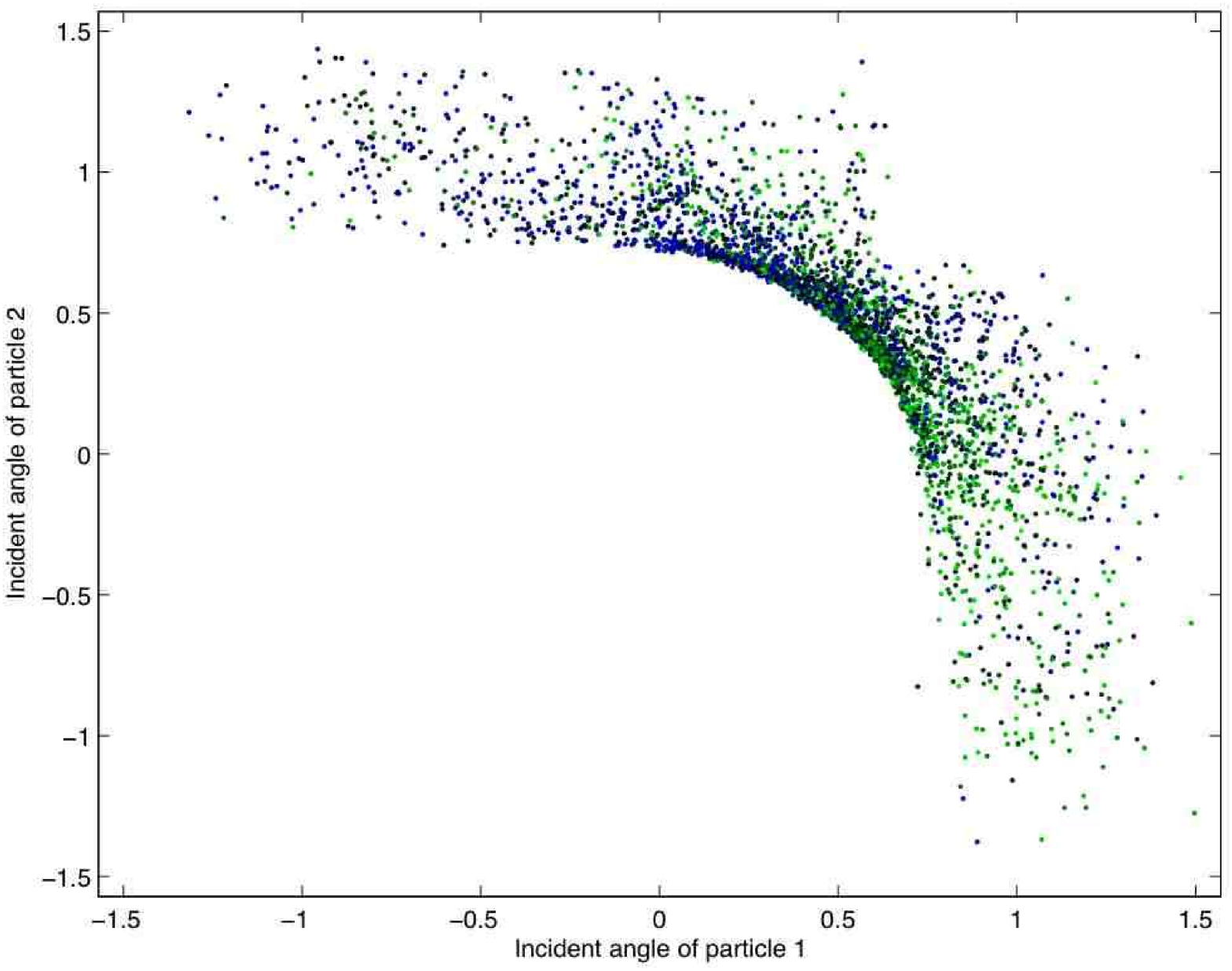}
                \hspace{.2 cm}
                (c)
                \includegraphics[width=0.2\textwidth]{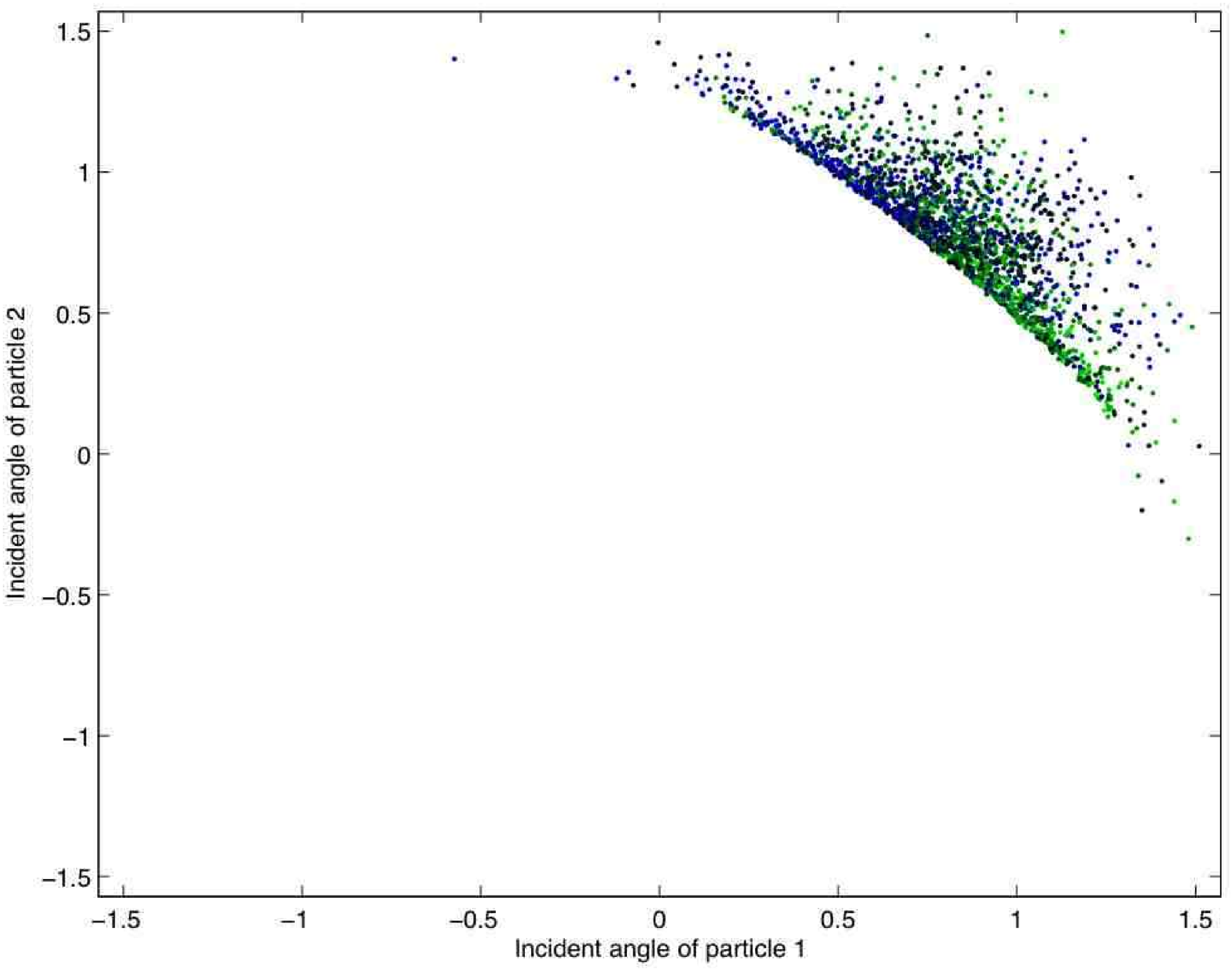}
                \hspace{.2 cm}
                (d)
                \includegraphics[width=0.2\textwidth]{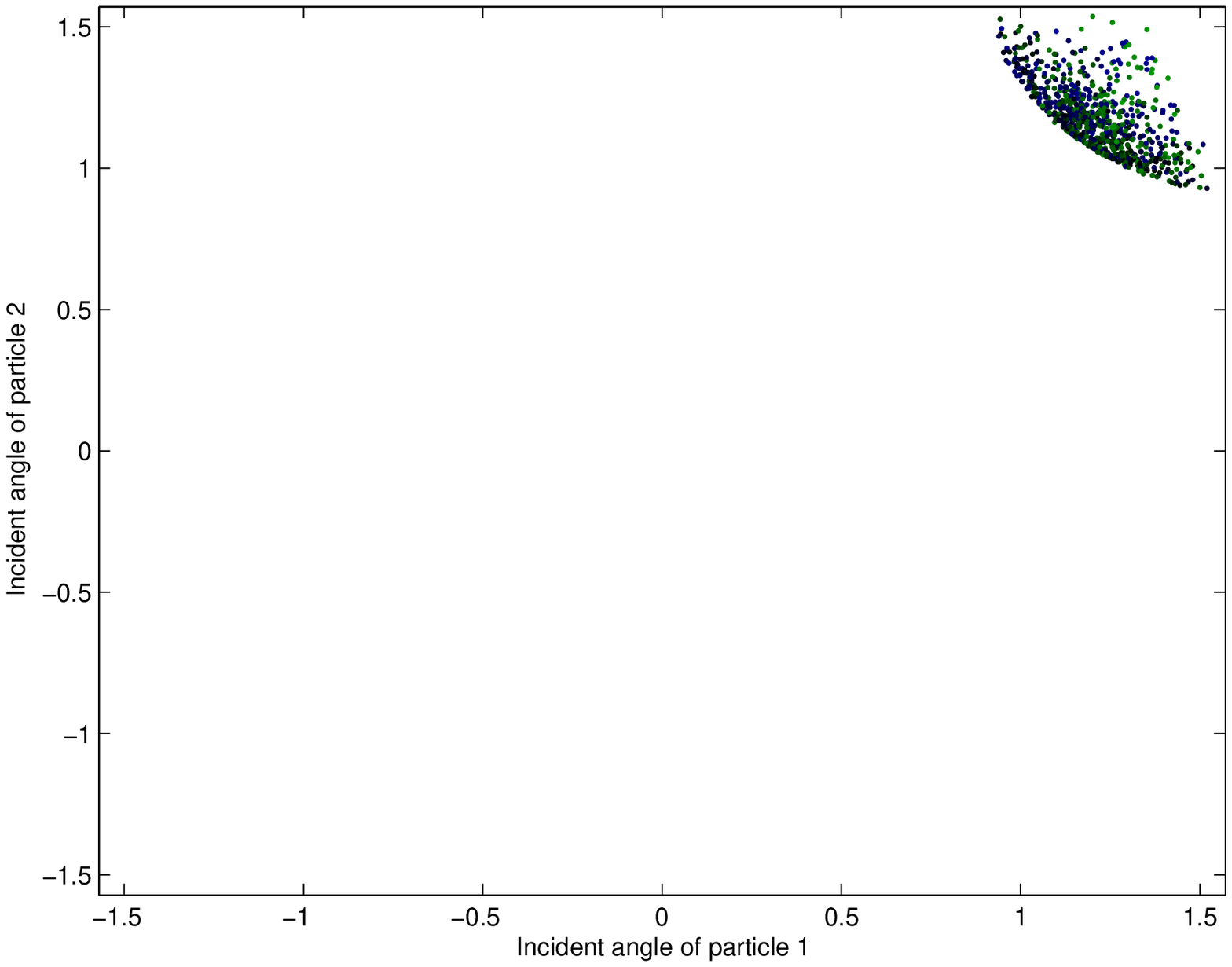}}
                \centerline{
                (e)
                \includegraphics[width=0.25\textwidth]{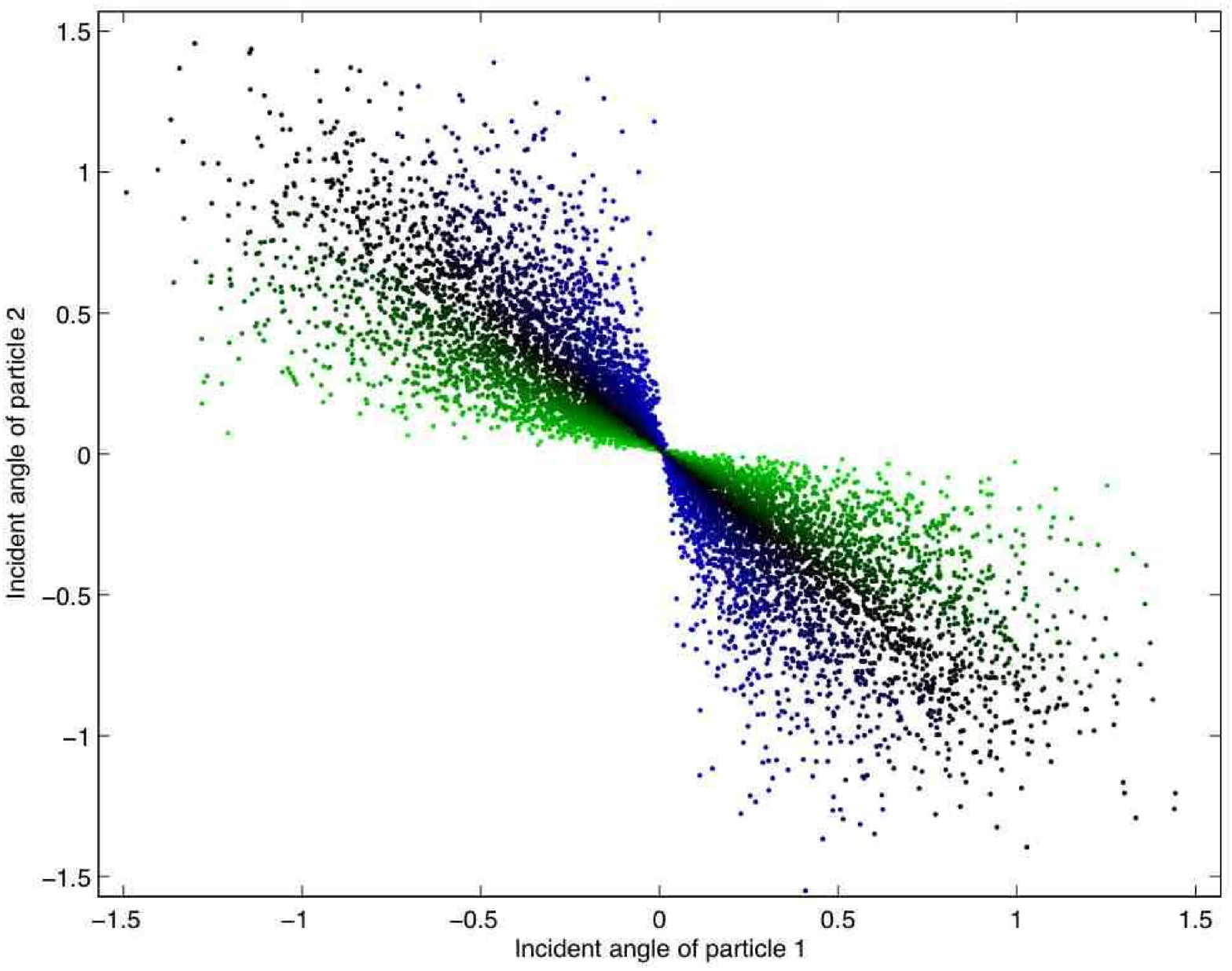}
                \hspace{.2 cm}
                (f)
                \includegraphics[width=0.25\textwidth]{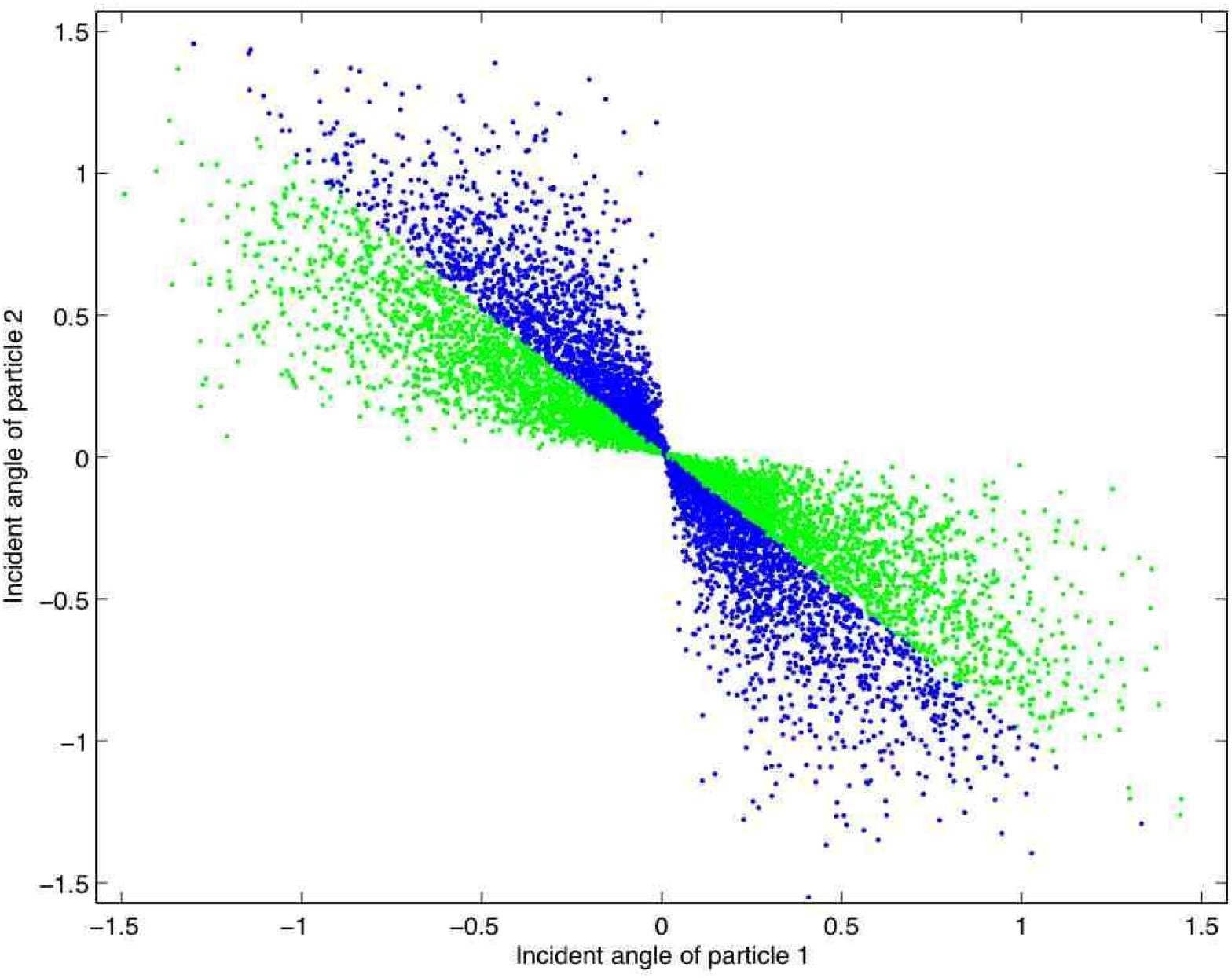}
                \hspace{.2 cm}
                (g)
                \includegraphics[width=0.25\textwidth]{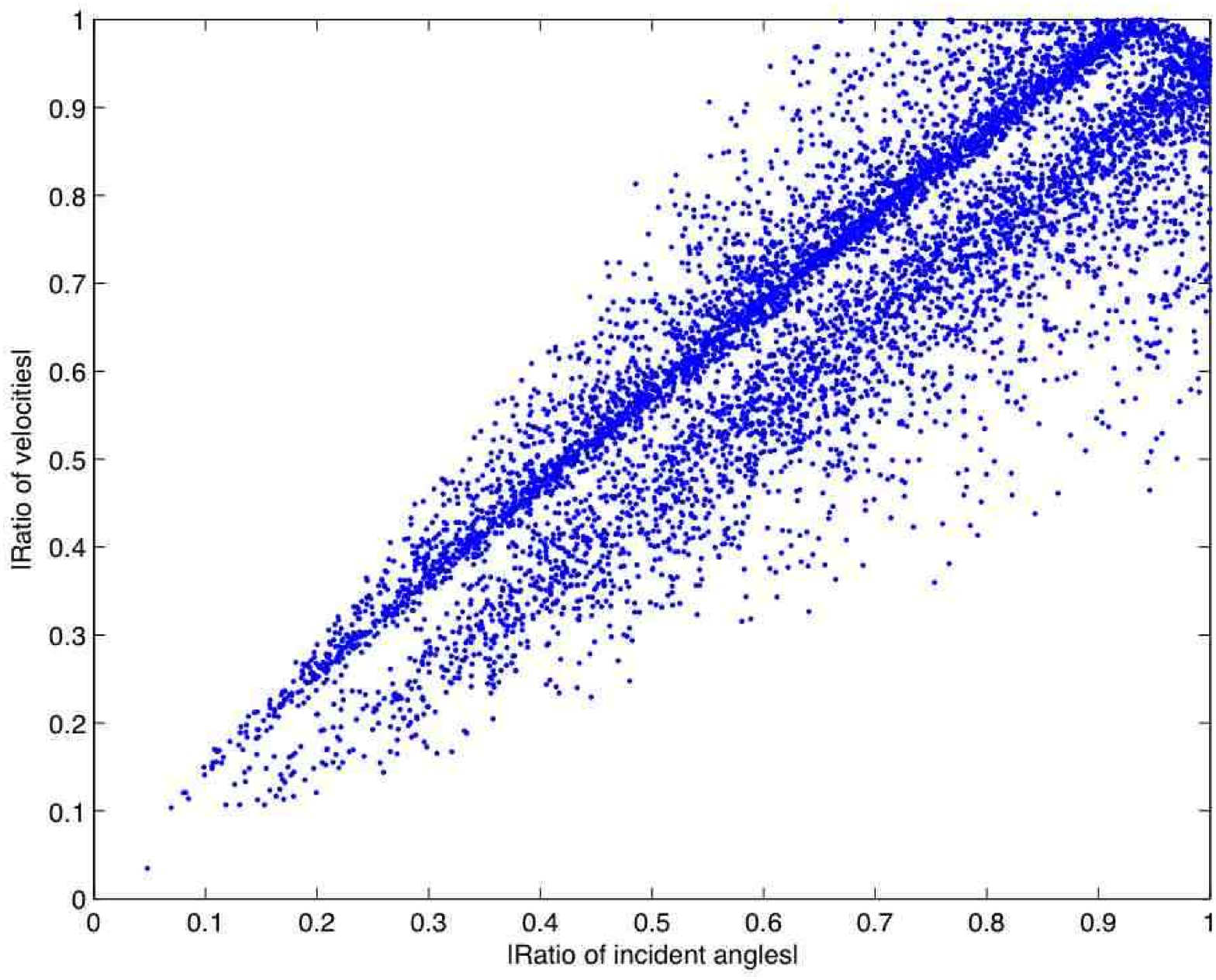}}   
                \caption{Two particles of radius $1/10$ in a circular billiard of unit radius starting in nearly the same or nearly opposite directions.  For each initial condition except that in panel (d), we numerically simulate the collisions between the two particles for 50000 total collisions.  (In all of our simulations, we use the terminology `total collisions' to indicate all the collisions between two particles and between either particle and the boundary.)  We run our simulation through 500000 such collisions in (d) to confirm the accuracy of the result.  Each point represents one collision between the two particles.  The horizontal axes depict the incident angle of particle 1, and the vertical axes depict the incident angle of particle 2.  Both particles have an initial speed of 1.  The first particle in all panels has initial horizontal coordinate $x_1(0) = 0.3$, and the second particle always has initial horizontal coordinate $x_2(0) = -0.3$.  In all cases, both particles have the same initial vertical coordinate $y(0)$.  The colors are defined as in Fig.~\ref{special}.  Panels (a)--(d) depict the results for particles with nearly the same initial angles [$\theta_1(0) = 0$, $\theta_2(0) = 0.1$], and panel (e) depicts the result for particles with nearly opposite initial angles [$\theta_1(0) = \pi$, $\theta_2(0) = 0.1$].  (We obtain the same results for smaller initial angle differences.)  For the case of nearly opposite angles, the dynamics are the same for different $y(0)$, so we only show one plot.  Panels (a) and (e) have $y(0) = 0.2$, (b) has $y(0) = 0.4$, (c) has $y(0) = 0.6$, and (d) has $y(0) = 0.8$.  (Color online) Panel (f) shows the same simulation as panel (e), except all collisions with $v_1 > v_2$ are shown in blue (dark gray) and all collisions with $v_1 < v_2$ are shown in green (light gray).  Panel (g), also for this simulation, shows a plot of the magnitude of the angle ratio versus the magnitude of the velocity ratio, indicating that they are approximately equal for particles with almost opposite initial angles.} \label{same}
\end{figure}

Now consider a pair of particles, again confined in circular containers, with nearly the same and nearly opposite initial velocities, so that we are perturbing from periodic orbits in which the particles collide against each other along the same sequence of chords of the circle.  We show plots of the incident collision angle of one particle versus the other for the former case in Fig.~\ref{same}a--d, with slightly different initial $y$-coordinate values in each plot, and for the latter case in Fig.~\ref{same}e.  (Given the same initial angles, one obtains essentially the same plots if the other initial conditions are changed.)  Observe the change in dynamics from panel (a) to panel (d).  We show only one panel for the case of nearly opposite velocities because the dynamics do not change with different $y(0)$.  The mean incident angle against the boundary (averaging over both particles) is nonzero in (a)--(d) but zero in (e).  Observe that the region in (d) is convex and does not develop any ``wings"  like the ones in panels (a)--(c), so that the ranges of incidence angles of the particles are very small for sufficiently large offset $y(0)$.  Using the same color-coding as in Fig.~\ref{special}, we see in the case of particles with nearly opposite initial velocities that if the incident angle of particle 1 is greater than that of particle 2, then particle 1 has a lower speed than particle 2, as the line indicating equal incident angles sharply divides the figure into four monochromatic quadrants (observe the symmetry between the two particles). This is demonstrated in panel (f), in which all collisions with $v_1/v_2 > 1$ are shown in blue (dark gray) and all collisions with $v_2/v_1 < 1$ are shown in green (light gray).  Panel (g) shows that in this case, the magnitude of the angle ratio between the two particles is approximately equal to that of the velocity ratio, so that one needs less information to effectively describe the state of the system than would otherwise be necessary.

\subsection{Two Particles in Geometries Corresponding to Chaotic Billiards}

\begin{figure}
                \centerline{
                (a)
                \includegraphics[width=0.25\textwidth]{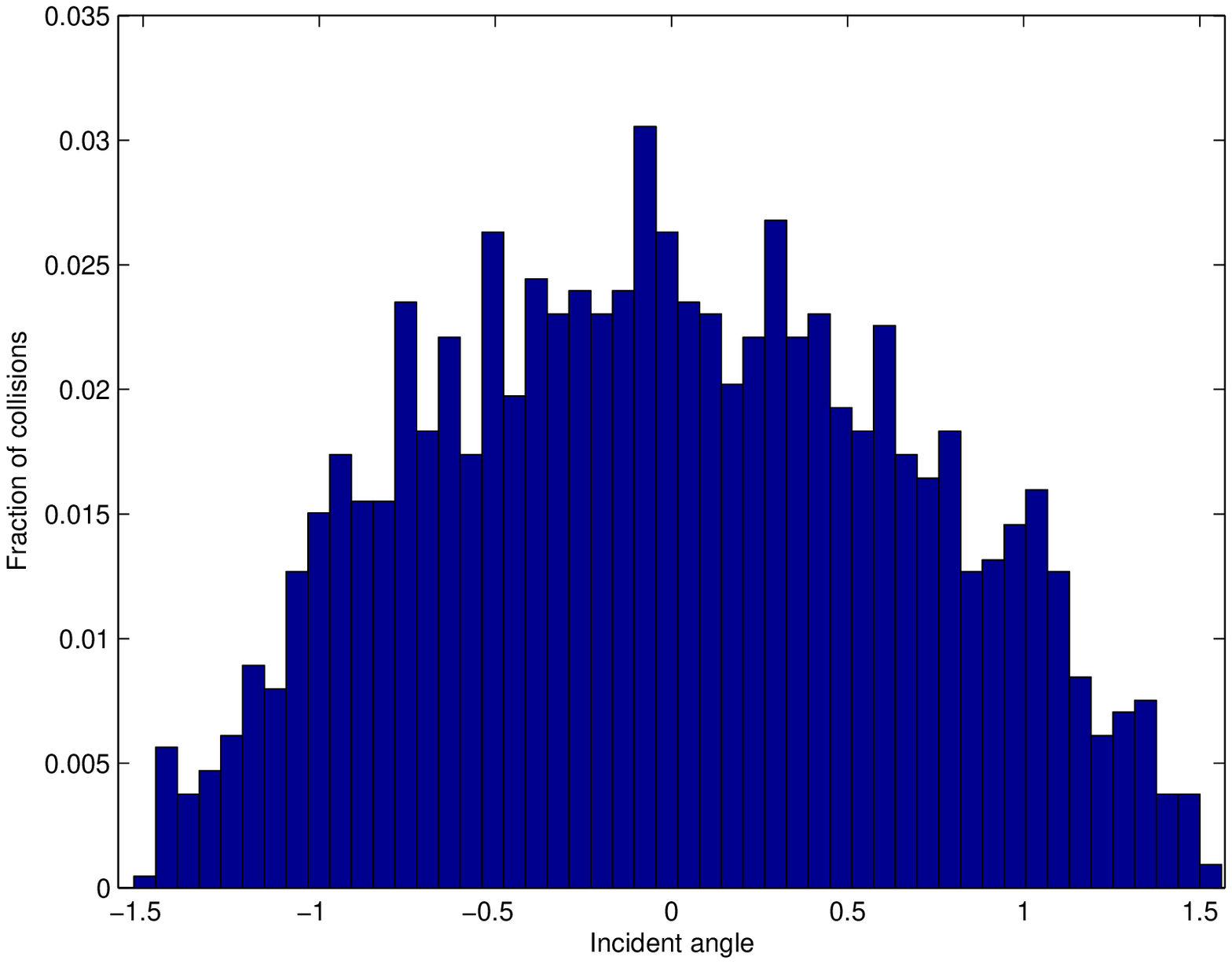}
                \hspace{.2 cm}
                (b)
                \includegraphics[width=0.25\textwidth]{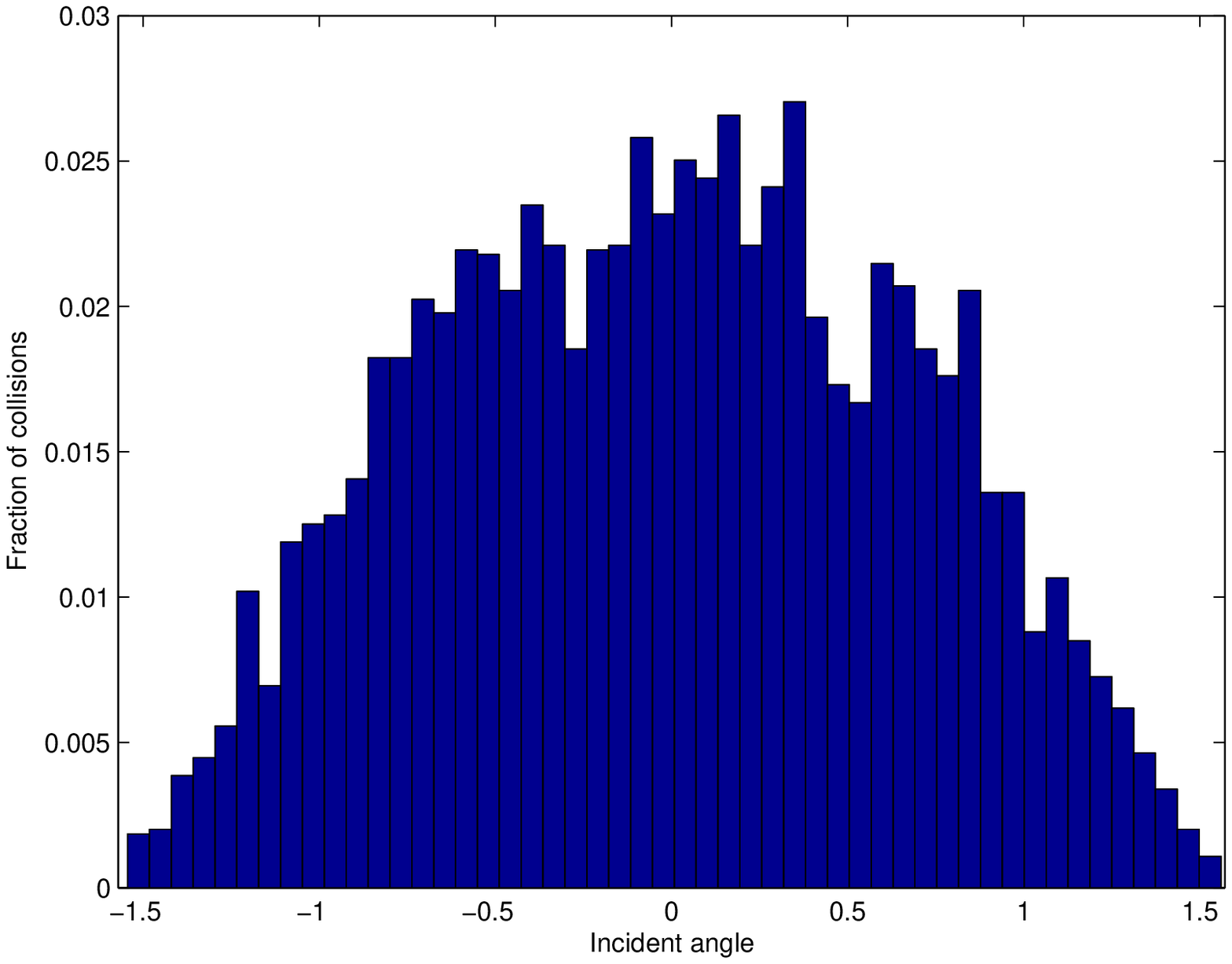}
                \hspace{.2 cm}
                (c)
                \includegraphics[width = 0.25\textwidth]{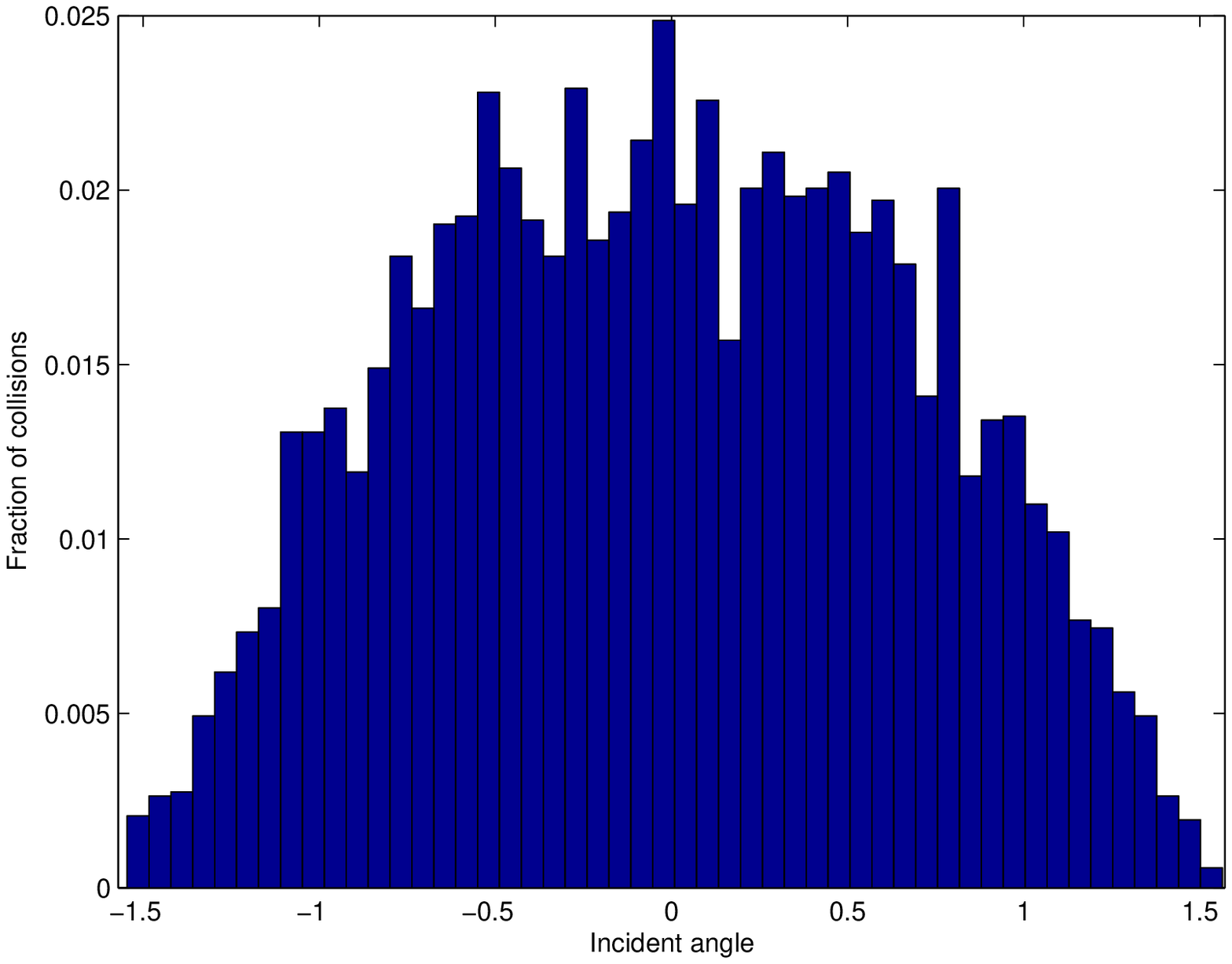}
                }

                \caption{Distribution of the incident angle of (several thousand) particle--boundary collisions for two hard balls of radius $1/4$ confined in (a,b) diamond containers composed of arcs of radius-1 circles and (c) a circular mushroom container.  The diamond in (a) has tangencies, as it consists of four quarter-circle arcs.  The diamond in (b) does not have tangencies, as it consists of four arcs that each constitute $1/8$th of a circle.}
\label{diamond}

\end{figure}

Diamond billiards are dispersing with boundaries consisting of arcs of four circles which cross at the vertices of a square.  We studied two-particle diamond billiards both with and without tangencies between the circular arcs.  In Fig.~\ref{diamond}ab, we show the distributions of their incident angles for particle--boundary collisions.  They differ markedly from those observed for two hard balls in circular containers (see Fig.~\ref{circle}).  In particular, the mean incident collision angle is $0$ for two-particle diamond billiards but nonzero for two-particle circular billiards.  In the diamond containers, the distributions fall off from the central peak to $0$ at about $\pm \pi/2$.  The distributions for two-particle diamond billiards are also wider than they are for circular ones.  Also unlike the case of circles, we observe here a higher density of collisions near the corners, so that the Poincar\'e map includes a clustering of points near the vertical lines corresponding to particle collisions with these corners.

\subsection{Two Particles in Geometries Corresponding to Billiards with Mixed Dynamics}

To examine two-particle billiards whose one-particle counterparts have mixed dynamics, we again consider containers shaped like mushrooms.  In Fig.~{\ref{diamond}c, we show the distribution of incident angles for particle--boundary collisions in the case of a circular mushroom.  This distribution is reminiscent of that for the diamond geometries in Fig.~\ref{diamond}ab.  In these two-particle simulations, we examined initial conditions with both particles in integrable regions of the corresponding one-particle billiard, both particles in chaotic regions, and one particle in each of these types of regions.  In all cases, we obtain the same results as with the diamond geometry (which, again, has a fully chaotic billiard limit).  We also observe these same dynamics for elliptical mushrooms with very eccentric caps.  (We considered examples with the stem attached to the foci, so that only one family of caustics is present in the one-particle billiard.)  If islands are present, they are probably very small in size.




\subsection{Three- and Four-Particle Billiards}

Because KAM tori divide phase space in two dimensions but not in higher dimensions, it is crucial to consider containers that confine three or more particles.  One may thus expect some differences in the dynamics when one considers more than two particles.  

To see that signatures of integrability remain even with additional particles, we considered three hard disks of radius $0.0816$ in a circle of unit radius.  (The area occupied by the three particles in these simulations is almost the same as that occupied by the radius $1/10$ particles in the corresponding two-particle simulations.)  In particular, we investigated initial conditions in which all three particles are collinear and initially traverse almost the same chord of the circle (i.e., their initial angles are almost the same).  As in the two-particle case, one of the balls has a small angular displacement (of $0.1$) with respect to the others and the common $y$-coordinate is varied exactly as for the two-particle simulations.  As shown in Fig.~\ref{threeball}, we again observe a non-zero mean in the distribution of incident angles for particle--boundary collisions.  (For a given simulation, the distribution for each of the individual particles is roughly the same, so we show the distributions for all three particles collectively.)  

We also observe a non-zero mean in the distribution of incident angles in a similar numerical experiment with four hard balls in a circle.  Our observations are summarized in Table \ref{tab2}.

\begin{figure}
                \centerline{
                (a)
                \includegraphics[width=0.2\textwidth]{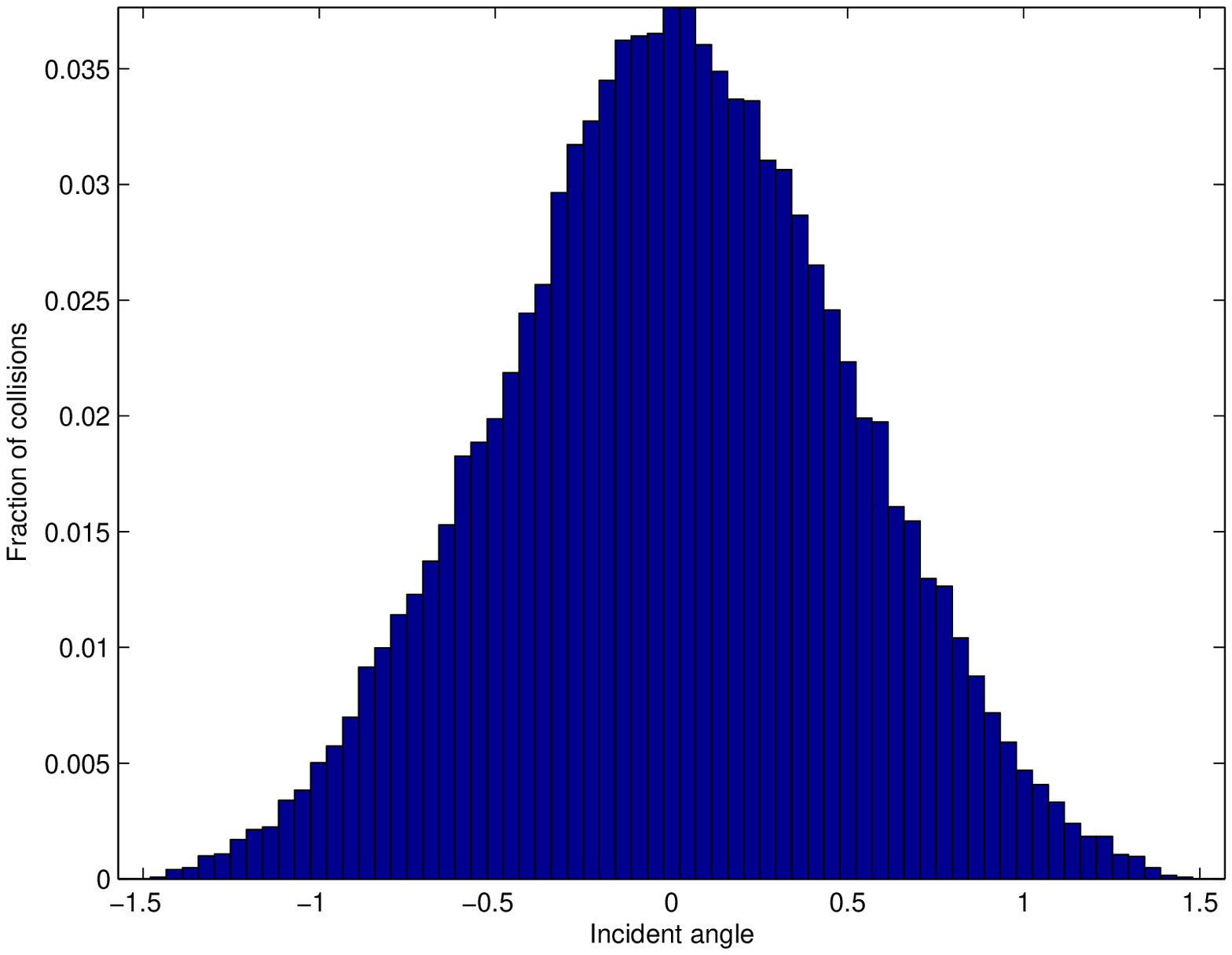}
                \hspace{.2 cm}
                (b)
                \includegraphics[width=0.2\textwidth]{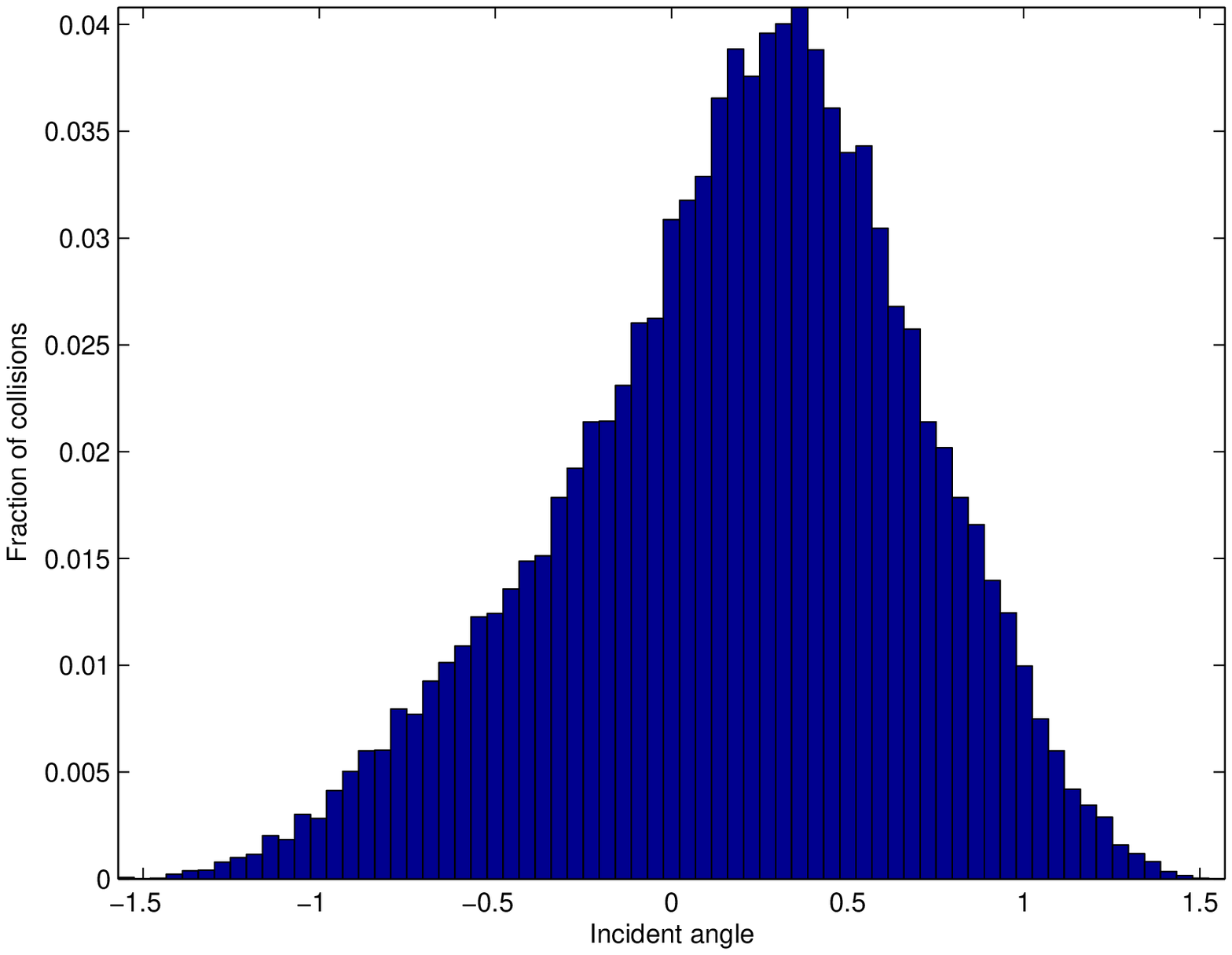}
                \hspace{.2 cm}
                (c)
                \includegraphics[width=0.2\textwidth]{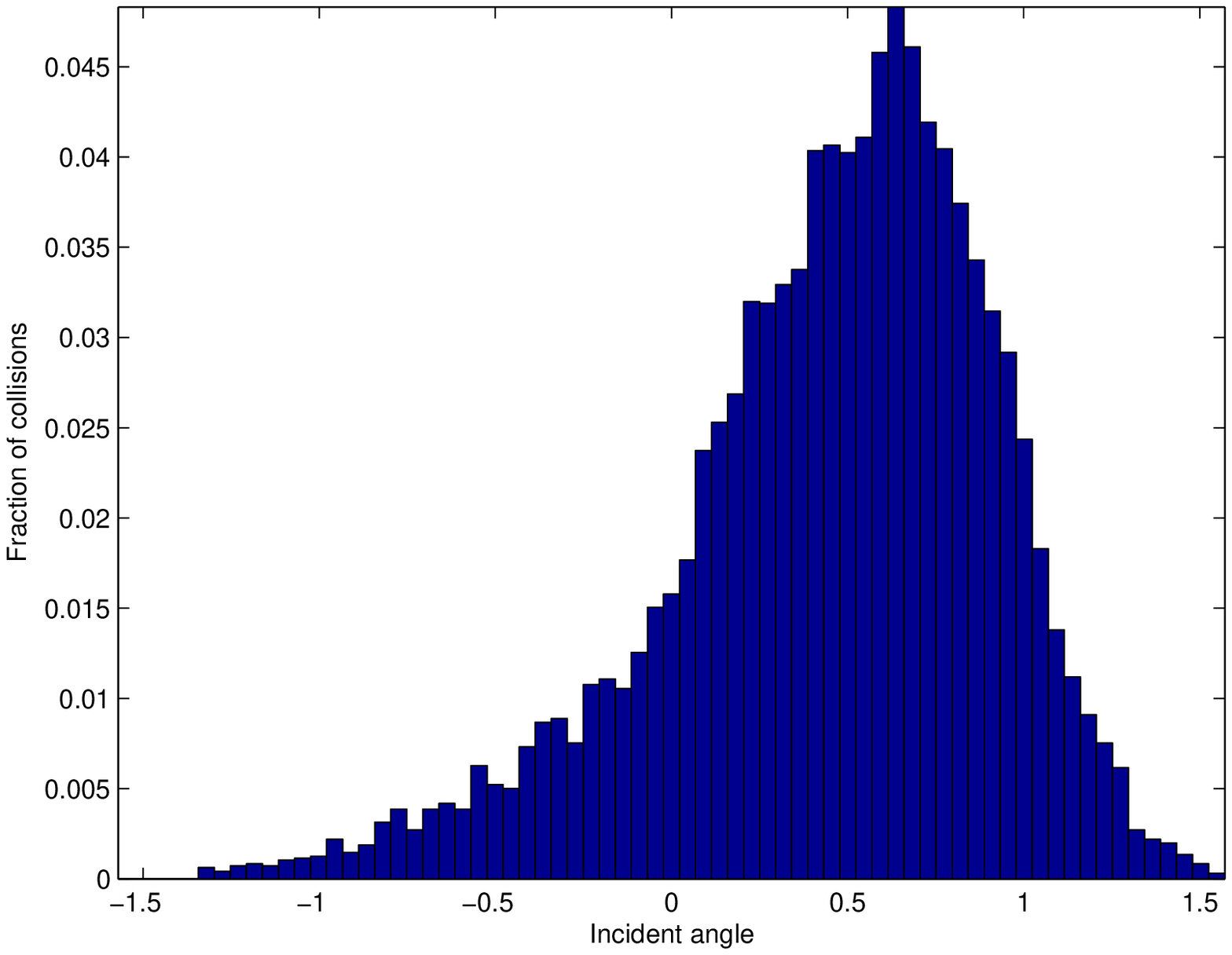}
                \hspace{.2 cm}
                (d)
                 \includegraphics[width=0.2\textwidth]{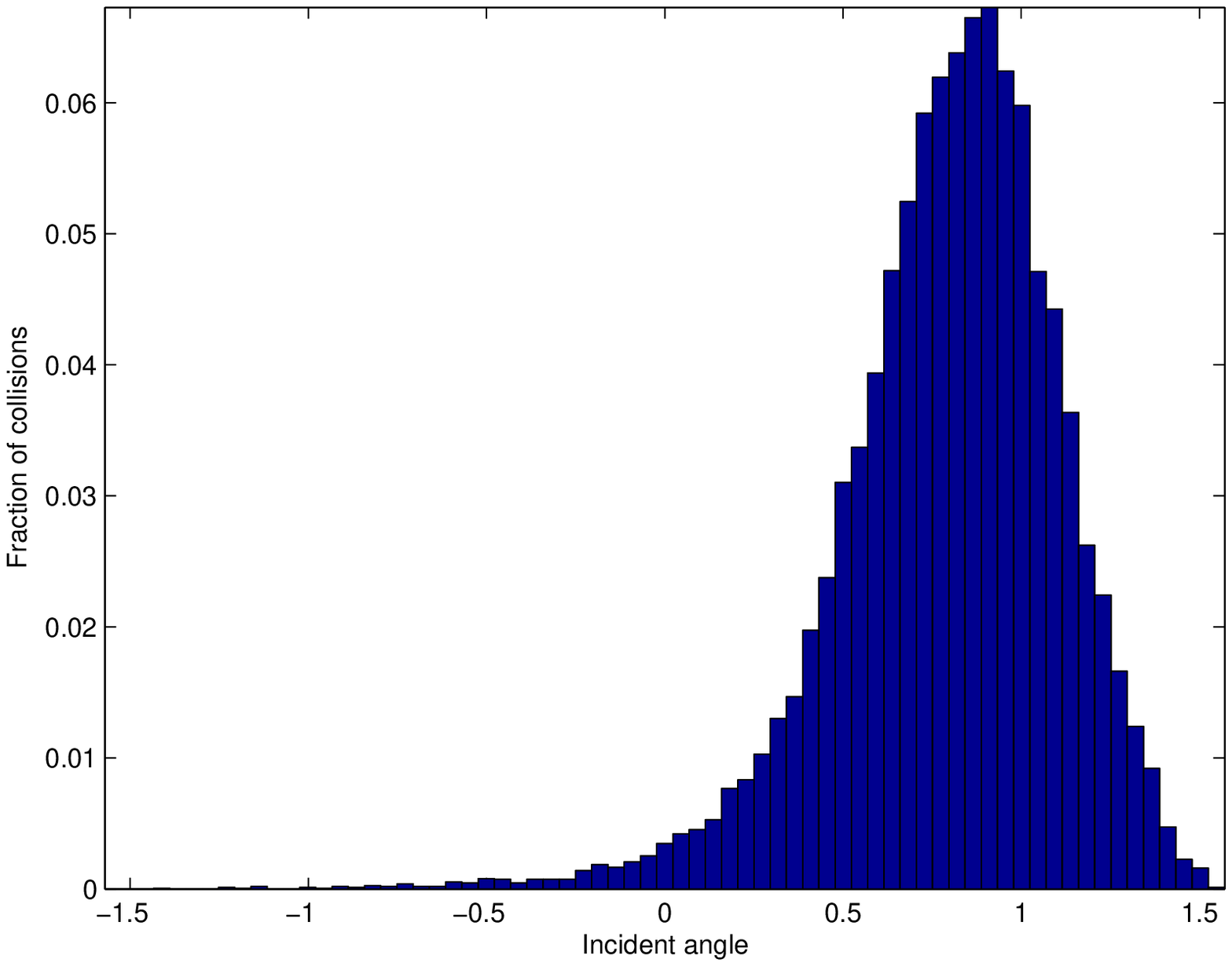}}

                \caption{Distribution of incident angle for particle--boundary collisions for three identical particles of radius $.0816$ confined in a circle of radius 1.  In each case (covering 100000 total collisions), we initialize all three particles with unit speed and the same $y$ coordinate (which is varied in the simulations).  For each simulation, the initial horizontal coordinates of the three particles are $x_1(0) = 0.3$, $x_2(0) = 0$, and $x_3(0) = -0.3$.  The initial angles are $\theta_1(0) = \theta_3(0) = 0$ and $\theta_2(0) = 0.1$.  We list the observed means and standard deviations in Table \ref{tab2}.  The initial vertical coordinates are (a) $y(0) = 0$, (b) $y(0) = 0.2$, (c) $y(0) = 0.4$, and (d) $y(0) = 0.6$.}
\label{threeball}

\end{figure}

\begin{table}
\centerline{\begin{tabular}{|c|c|c|c|c|c|c|} \hline
$y(0)$ & $\mu_2$ & $\mu_3$ & $\mu_4$ & $\sigma_2$ & $\sigma_3$ & $\sigma_4$  \\ \hline
0    & 0.0083 & 0.0088 & -0.0085 & 0.4256 & 0.4863 & 0.5195 \\
0.2 & 0.2143 & 0.1893 & 0.1894  & 0.4788 & 0.4899 & 0.5189 \\
0.4 & 0.4829 & 0.4529 & 0.4443  & 0.4283 & 0.4627 & 0.4793 \\
0.6 & 0.8241 & 0.7957 & 0.7852  & 0.2485 & 0.3132 & 0.3490 \\ \hline
\end{tabular}}
\caption{Mean $\mu_j$ and standard deviation $\sigma_j$ for the incident angle of particle--boundary collisions for $j = 2, 3, 4$ hard balls in a circular container.  In all simulations (of 100000 total collisions), each particle has initial speed $1$ and initial $y$-coordinate $y(0)$.  The other initial conditions for $j = 2$ ($j = 3$) are shown in Fig.~\ref{same} (Fig.~\ref{threeball}).  For $j = 4$, we considered circular particles of radius $0.0707$.  For each simulation, the initial $x$-coordinates are $x_1(0) = 0.3$, $x_2(0) = 0.1$, $x_3(0) = -0.1$, and $x_4(0) = -0.3$.  The initial angles are $\theta_1(0) = \theta_3(0) = \theta_4(0) = 0$ and $\theta_2(0) = 0.1$.  Observe that the mean decreases and the standard deviation increases with the additional particles.}
\label{tab2}
\end{table}

\section{Conclusions}

In this paper, we examined the dynamics of one-particle and few-particle billiards.  These situations can be compared in the context of noninteracting versus interacting hard balls in a given container.  (The "billiard limit" discussed in the paper in which the particle size becomes smaller also provides a means for comparison.)  For example, two noninteracting hard balls in a circular container are integrable, as this system just consists of two integrable billiards.  We showed using numerical simulations that two interacting hard balls in a circular container bear signatures of integrability even though the system is chaotic and that its dynamics differs from that of two particles in a diamond container (which is chaotic even in the one-particle case).  Also, two noninteracting particles in a mushroom show mixed dynamics, whereas two interacting particles in a mushroom container seem to show the same dynamics as two interacting particles in a diamond.

Consequently, we have shown that the shape of a container confining a system of interacting particles {\it does} matter, despite the fact that this aspect of the dynamics in the behavior of a gas of particles seems to be essentially neglected in investigations of these systems. In fact, the container shape may be of comparable importance to that of the interaction potential. Although the importance of the shape of the container (the so-called ``billiard table") for systems of noninteracting particles is universally acknowledged, a ``silent consensus" still exists that if the interaction potential is nontrivial, then a system of interacting particles must be ergodic in the thermodynamic limit. We have shown in this paper, however, that this is not true for systems of a few interacting particles, which have thus far received little attention.  With the emergence of nanosystems and new experimental techniques, such systems are now among the most important ones for various applications \cite{atomoptics,cornell}.

Exactly solvable billiard systems demonstrating all three possible types of behavior (integrability, chaos, and mixed dynamics) motivate natural container shapes to study the dynamics of a few nontrivially interacting  particles. While the first small step has been made in this paper, most of the problems in this area remain completely open; we mention a few of them in passing.  First, it is well-known in statistical physics that the high-density limit is more regular than the low-density limit (where, for example, logarithmic terms appear in the asymptotics of transport coefficients \cite{leener}).  We have observed this in our numerical studies as well.  In fact, it seems to be a general phenomenon for which a theory should be developed without the ubiquitous assumption that the number of particles is a large parameter. As usual, the temporal asymptotics of both correlation functions and Poincar\'e recurrences should be investigated in various situations, including all three types of behavior in billiards (noninteracting particles) and for broad classes of potentials.  One of the key questions is what happens to the relative volume of KAM islands when the number of particles tends to infinity \cite{mush2}.  Again the ``silent consensus" (i.e., a belief without proofs or convincing demonstrations) is that this relative volume tends to zero. However, it has been demonstrated that this is at least not always true \cite{mush2}.  Hence, it is not at all unlikely that nonlinear dynamics will reveal some new surprises in further investigations of this phenomenon.

\section*{Acknowledgements}

SL and MAP acknowledge support provided by an NSF VIGRE grant awarded to the School of Mathematics at Georgia Tech.  The work of LB and SL was partially supported by NSF grant DMS 0140165.  LB also acknowledges the support of the Humboldt Foundation, SL acknowledges funding from a Georgia Tech PURA award, and MAP acknowledges support from the Gordon and Betty Moore Foundation through Caltech's Center for the Physics of Information.  We also thank Nir Davidson for useful conversations and the anonymous referee for suggestions that led to significant improvements in the paper.


\end{document}